\documentclass[12pt]{article}
\usepackage[cp1251]{inputenc}  
\usepackage[english]{babel}
\usepackage{amsmath,amssymb,epsfig,graphicx,doc,latexsym,amsfonts,bm,longtable}
\usepackage[bookmarksopen,colorlinks]{hyperref}
\usepackage[monochrome]{color}
 \hypersetup{colorlinks=true}

\begin{document}
\begin{center}
{\textbf{\LARGE{Renormalized non-modal theory of the kinetic drift
instability of plasma shear flows}}}

\bigskip
\bigskip

{\large{V.S. Mikhailenko$^{1}$}}, {\large{V.V. Mikhailenko$^{2}$,
K.N. Stepanov$^{2}$}}

\end{center}
\bigskip
$^{1}$V.N. Karazin Kharkov National University, 61108 Kharkov,
Ukraine
\\
$^{2}$University of Madeira, Largo do Municipio, 9000 Funchal,
Portugal
\\
$^{3}$National Science Center ``Kharkov Institute of Physics and
Technology'', 61108 Kharkov, Ukraine

\bigskip

\begin{abstract}
The linear and renormalized nonlinear kinetic theory of drift
instability of plasma shear flow across the magnetic field, which
has the Kelvin's method of shearing modes or so-called non-modal
approach as its foundation, is developed. The developed theory
proves that the time-dependent effect of the finite ion Larmor
radius is the key effect, which is responsible for the suppression
of drift turbulence in an inhomogeneous electric field. This effect
leads to the non-modal decrease of the frequency and growth rate of
the unstable drift perturbations with time. We find that turbulent
scattering of the ion gyrophase is the dominant effect, which
determines extremely rapid suppression of drift turbulence in shear
flow.
\\ 52.35.Ra  52.35.Kt
\end{abstract}

\section*{I.INTRODUCTION}
After seminal investigations of the regimes of the enhanced plasma
confinement\cite{Wagner}, the understanding the role of flow shears
in turbulent transport becomes one of the major issues of tokamak
plasmas physics\cite{burell97-1499}. The stabilizing effect of the
flow shears on various drift waves is recognized as one of the
essential elements in the formation of core and edge transport
barriers. A decisive step in understanding the role of flow shears
in the reducing of the turbulent transport and formation of
transport barriers has been made with the development of the
gyrokinetic theory for describing turbulence in plasma with large
shear flows \cite{Brizard}-\cite{Wang} and the development of
numerical codes\cite{Waltz}-\cite{Waltz2} able to solve the set of
gyrokinetic equations. In spite of the great progress in numerical
investigations of tokamak turbulence, there still remain some key
issues in analytical investigations of the turbulence in plasma
shear flows, which deserve further clarification. One of the most
challenging problem in the theory of plasma shear flows turbulence
is the development of analytical methods of the investigations of
the long-time evolution of the turbulence in shear flow governed by
Vlasov-Maxwell system. Applications of the gyrokinetic approach to
analytical investigation of the instabilities in plasma shear flows
has been made in series of papers \cite{Artun}-\cite{Maccio}, in
which the electrostatic ion-temperature-gradient mode was
considered. In all these papers, the spectral transform in time was
applied and perturbed electrostatic potential $\Phi$ was considered
in canonical modal form, $\Phi\sim \exp\left(-i\omega t\right)$.
That modal form, however, unsatisfactory represents the long-time
response of plasma shear flows across the magnetic field due to
space-dependent Doppler shift and stretching of waves pattern by
shear flows\cite{Mikhailenko-2000}. In fact, that approach gives
results which are valid only for times limited by the condition
$t\ll \left(V'_{0}\right)^{-1}$, where $V'_{0}$ is flow velocity
shear.

One of the most effective approaches to the analysis of the temporal
evolution of plasma turbulence in shear flows is method of shearing
modes or so-called non-modal approach. The essence of this approach,
which originally was developed by Lord Kelvin\cite{Kelvin} for fluid
flows with a homogeneous velocity shear, consists in transforming
the independent spatial variables from the laboratory frame to a
frame convected with shear flow and studying the temporal evolution
of the spatial Fourier modes of perturbation without any spectral
expansions in time (\cite{Mikhailenko-2000}, \cite{Mikhailenko2010}
and references therein). The transformation to the coordinates
convected with shear flow eliminates the explicit spatial dependence
related to shear flow from the convective derivative in governing
fluid equations. This transformation not only simplifies governing
equations, but is also principally indispensable. The temporal
evolution of a separate spatial Fourier harmonic with a definite
wave number can only be analyzed with convective coordinates; it is
in contrast to the laboratory set of reference, in which spatial
Fourier harmonics are coupled due to velocity shear.

Kinetic effects, such as finite Larmor radius effects, Landau and
cyclotron damping and the numerous resulting kinetic instabilities,
which are naturally not involved in the fluid description of plasma
shear flows, require the development of a non-modal kinetic
description of plasma shear flows. Note also, that because of the
shearing of perturbations in shear flow, the component of the wave
number along the direction of the velocity shear experiences secular
growth with time\cite{Mikhailenko-2000},\cite{Mikhailenko2010}, and
therefore results obtained on the base of the fluid description are
valid only for finite times at which the condition
$k_{\perp}\left(t\right)\rho_{i}\ll 1$, ($k_{\perp}$ is the
component of the wave number across the magnetic field and
$\rho_{i}$ is thermal ion Larmor radius) for initially long
wavelength perturbations with
$k_{\perp}\left(t=t_{0}\right)\rho_{i}\ll 1$ holds. All this
requires the development of the non-modal kinetic theory, which can
properly describe the long-time evolution of the perturbations in
shear flow. Motivated by these requirements, we have undertaken to
work out the kinetic theory of plasma shear flows, which has
Kelvin's method of shearing modes, or the so-called non-modal
approach, as its foundation. The present paper focuses on the
development of basic equations for linear and renormalized nonlinear
non-modal kinetic theory for electrostatic perturbations of shear
flow with homogeneous velocity shear and their solutions for kinetic
drift instability of plasma shear flow. As a first attempt of such
analytical investigation, we consider here the case of the
homogeneous shearless magnetic field. In Section II, the governing
linear integral equation for electrostatic potential is derived. In
Section III, the solution of that equation for drift kinetic
instability, which displays the non-modal evolution with time of the
electrostatic potential in shear flow, is obtained. The nonlinear
integral equation for electrostatic potential, which accounted for
the scattering of ions by shearing modes with random phases in
plasma shear flow, is obtained in Section IV. The solution of that
equation, which describes the suppression of kinetic drift
instability in shear flow is obtained in Section V. A summary of the
work is given in Conclusions, Section IV.

\section*{II.VLASOV--POISSON SYSTEM OF EQUATIONS \newline IN SHEARED COORDINATES}
Our theory is based on the Vlasov equations,
\begin{eqnarray}
&\displaystyle \frac{\partial F_{\alpha}}{\partial
t}+\mathbf{v}\frac{\partial F_{\alpha}}
{\partial\mathbf{r}}+\frac{e_{\alpha}}{m_{\alpha}}\left(\mathbf{E}_{0}
\left(\mathbf{r}\right)
+\frac{1}{c}\left[\mathbf{v}\times\mathbf{B}\right]
-\bigtriangledown
\varphi\left(\mathbf{r},t\right)\right)\frac{\partial
F_{\alpha}}{\partial\mathbf{v}}=0,\label{1}
\end{eqnarray}
for velocity distribution function $F_{\alpha}$ of $\alpha$ species
($\alpha=i$ for ions and $\alpha=e$ for electrons) in inhomogeneous
electric field $\mathbf{E}_{0}\left(\mathbf{r}\right)$, directed
across the homogeneous magnetic field $\mathbf{B}$, and Poisson
equation for the perturbed electrostatic potential
$\varphi\left(\mathbf{r},t\right)$,
\begin{eqnarray}
&\displaystyle \vartriangle \varphi\left(\mathbf{r},t\right)=
-4\pi\sum_{\alpha=i,e} e_{\alpha}\int f_{\alpha}\left(\mathbf{v},
\mathbf{r}, t \right)d\textbf {v}_{\alpha}, \label{2}
\end{eqnarray}
where  $f_{\alpha}$ is the perturbation of the equilibrium
distribution function $F_{0\alpha}$,
$F_{\alpha}=F_{0\alpha}+f_{\alpha}$. The kinetic theory for shear
flows is developed as a rule\cite{ Hahm, Artun} in a frame that is
shifted by $\mathbf{V}_{0}\left(x\right)$ in velocity space, but is
unchanged in configuration space, leaving the inhomogeneous
convective terms in Vlasov equation. The starting point of the
derivation of the basic equations of the nonmodal kinetic theory is
the transformation of Vlasov-Poisson system to convective (sheared)
coordinates in velocity and configuration spaces.

We transform in Eq.(\ref{1}) the variables $\left( t, \textbf{r},
\textbf{v}\right) $ onto new spatial variables $\left(
t_{c},\textbf{r}_{\alpha},\textbf{v}_{\alpha}\right) $, connected by
the relations
\begin{eqnarray}
&\displaystyle
t=t_{c},\qquad\mathbf{v}=\mathbf{v}_{\alpha}+\mathbf{U}_{\alpha}
\left(\mathbf{r}_{\alpha}, t_{c}\right), \qquad
\mathbf{r}=\mathbf{r}_{\alpha}+\int\limits^{t_{c}}_{t_{(0)}}\mathbf{U}_{\alpha}
\left(\mathbf{r}_{\alpha}, t_{1c}\right)dt_{1c},\label{3}
\end{eqnarray}
or
\begin{eqnarray}
&\displaystyle
\mathbf{v}_{\alpha}=\mathbf{v}-\mathbf{V}_{\alpha}\left(\mathbf{r}_{\alpha},
t\right), \qquad
\mathbf{r}_{\alpha}=\mathbf{r}-\int\limits^{t}_{t_{(0)}}\mathbf{V}_{\alpha}\left(\mathbf{r},
t_{1}\right)dt_{1},\label{4}
\end{eqnarray}
with set of reference moving with velocity
$\mathbf{V}_{\alpha}\left(\textbf{r}, t \right) =
\mathbf{U}_{\alpha}\left(\textbf{r}_{\alpha}, t_{c} \right)$. In
Eqs.(\ref{3}),(\ref{4}) $t_{(0)}$ denotes the time at which shear
flow emerges. For the transformation of Eq.(\ref{1}) to new
variables we use the relations, which follows from
Eqs.(\ref{3}),(\ref{4}),
\begin{eqnarray}
&\displaystyle \frac{\partial}{\partial t}=\frac{\partial}{\partial
t_{c}}-\frac{\partial V_{i\alpha} \left( \mathbf{r}, t\right)
}{\partial t}\frac{\partial}{\partial v_{i\alpha}}-V_{i\alpha}
\left( \mathbf{r}, t\right)\frac{\partial}{\partial
r_{i\alpha}},\label{5} \\
&\displaystyle \frac{\partial}{\partial
v_{i}}=\frac{\partial}{\partial v_{i\alpha}},\qquad \frac{\partial
v_{i\alpha}}{\partial r_{j} }=-\frac{\partial V_{i\alpha}}{\partial
r_{j}}, \qquad \frac{\partial r_{i\alpha}}{\partial r_{j}
}=\delta_{ij}-\int\limits^{t}_{t_{(0)}} \frac{\partial
V_{i\alpha}\left( \mathbf{r}, t_{1}\right)} {\partial
r_{j}}dt_{1},\label{6}
\end{eqnarray}
With these relations Eq.(\ref{1}) takes the form
\begin{eqnarray}
&\displaystyle \frac{\partial F\left(
t_{c},\mathbf{r}_{\alpha},\mathbf{v}_{\alpha}\right)} {\partial
t_{c}}+v_{i\alpha}\frac{\partial F\left(
t_{c},\mathbf{r}_{\alpha},\mathbf{v}_{\alpha}\right)} {\partial
r_{i\alpha}}-\left(v_{i\alpha}+ U_{i\alpha}\left(
\mathbf{r}_{\alpha}, t_{c}\right)\right)
\int\limits^{t}_{t_{(0)}}\frac{\partial V_{j\alpha}\left(
\mathbf{r}, t_{1}\right)} {\partial r_{i}}dt_{1}\frac{\partial
F\left( t_{c},\mathbf{r}_{\alpha},\mathbf{v}_{\alpha}\right)}
{\partial r_{j\alpha}}\nonumber \\ &\displaystyle
-v_{i\alpha}\frac{\partial V_{j\alpha} \left( \mathbf{r},
t_{1}\right)}{\partial r_{i}}\frac{\partial F\left(
t_{c},\mathbf{r}_{\alpha}, \mathbf{v}_{\alpha}\right)}{\partial
v_{j\alpha}}+\frac{e_{\alpha}}{m_{\alpha}c}\left[\mathbf{v}_{\alpha}
\times\mathbf{B}\right]\frac{\partial
F_{\alpha}}{\partial\mathbf{v}_{\alpha}}\nonumber \\ &\displaystyle
-\left\{ \left[ \frac{d V_{i\alpha}}{d
t}-\frac{e_{\alpha}}{m_{\alpha}}\left( E_{0i}\left(\mathbf{r}\right)
+\frac{1}{c}\left[\mathbf{V}_{\alpha}\times\mathbf{B}\right]_{i}
\right) \right] -\frac{e_{\alpha}}{m_{\alpha}}
\frac{\partial\varphi\left(\mathbf{r},t\right)}{\partial
r_{i}}\right\}\frac{\partial F\left( t_{c},
\mathbf{r}_{\alpha},\mathbf{v}_{\alpha}\right)}{\partial
v_{i\alpha}} =0,\label{7}
\end{eqnarray}
where $dV_{i\alpha}/dt=\partial V_{i\alpha}/\partial
t+V_{j\alpha}\left( \partial V_{i\alpha}/
\partial r_{j}\right)$. When the velocity $V_{i\alpha}\left( \mathbf{r}, t\right)$ is such as
it vanishes the square brackets on last line of Eq.(\ref{7}), Vlasov
equation for shear flows, for which $U_{i\alpha}\left(
\textbf{r}_{\alpha}, t_{c}\right) \partial V_{j\alpha} \left(
\textbf{r}, t\right)/\partial r_{i}\equiv 0$, contains only velocity
shear parameter, $V'_{\alpha}$, (instead of $V_{\alpha}(\textbf{r})$
with laboratory frame variables),
\begin{eqnarray}
&\displaystyle \frac{\partial F_{\alpha}}{\partial t}+v_{\alpha
x}\frac{\partial F_{\alpha}}{\partial x} -\left(v_{\alpha
y}-v_{\alpha x}V'_{\alpha}t \right) \frac{\partial
F_{\alpha}}{\partial y} +\omega_{c\alpha} v_{\alpha y}
\frac{\partial F_{\alpha}}{\partial v_{\alpha
x}}-\left(\omega_{c\alpha}+V'_{\alpha} \right) v_{\alpha
x}\frac{\partial F_{\alpha}}{\partial v_{\alpha y}} \nonumber
\\  &\displaystyle
-\frac{e_{\alpha}}{m_{\alpha}}\left(\frac{\partial \varphi}{\partial
x} -V'_{\alpha}t\frac{\partial \varphi}{\partial y} \right)
\frac{\partial F_{\alpha}}{\partial v_{\alpha x}}+v_{\alpha
z}\frac{\partial F_{\alpha}}{\partial z_{\alpha}}
-\frac{e_{\alpha}}{m_{\alpha}} \frac{\partial \varphi}{\partial y}
\frac{\partial F_{\alpha}}{\partial v_{\alpha y}}
-\frac{e_{\alpha}}{m_{\alpha}} \frac{\partial \varphi}{\partial
z_{\alpha}} \frac{\partial F_{\alpha}}{\partial v_{\alpha
z}}=0.\label{8}
\end{eqnarray}

In this paper we consider stationary plasma shear flows across the
magnetic field with homogeneous velocity shear, for which
\begin{eqnarray}
&\displaystyle \mathbf{V}_{\alpha}=\mathbf{V}_{0}
=-\frac{c}{B}\frac{dE_{0}}{dx}x\mathbf{e}_{y}
=\frac{dV_{0}}{dx}x\mathbf{e}_{y},\label{9}
\end{eqnarray}
with $\mathbf{E}_{0} \left(\mathbf{r}\right)=\left(dE_{0}/dx\right)x
\mathbf{e}_{x}$ and $dE_{0}/dx=const$. For this case the
transformations (\ref{3}) have a form
\begin{eqnarray}
&\displaystyle t=t_{c}, \qquad v_{x}=v_{\alpha x},\qquad
v_{y}=v_{\alpha y}+V'_{0}x_{\alpha},\qquad v_{z}
=v_{z\alpha},\label{10}
\\ &\displaystyle x=x_{\alpha},\qquad
y=y_{\alpha}+V'_{0}x_{\alpha}t_{c}, \qquad z=z_{\alpha}\label{11}
\end{eqnarray}
where $V'_{0}=dV_{0}/dx$, and $t_{(0)}=0$ was assumed.

With leading center coordinates,
\begin{eqnarray}
&\displaystyle
X_{\alpha}=x_{\alpha}+\frac{v_{\bot}}{\sqrt{\mu_{\alpha}}\omega_{c\alpha}}\sin
\left( \phi_{1}-\sqrt{\mu_{\alpha}}\omega_{c\alpha}t\right),
\nonumber \\ &\displaystyle
Y_{\alpha}=y_{\alpha}-\frac{v_{\bot}}{\mu_{\alpha}\omega_{c\alpha}}\cos
\left( \phi_{1}-\sqrt{\mu_{\alpha}}\omega_{c\alpha}t\right)
-V'_{0}t\; \left(X_{\alpha}-x_{\alpha}\right), \qquad
z_{1}=z-v_{z}t,\label{12}
\end{eqnarray}
and velocity space coordinates (see, also Ref.\cite{Shaing}),
\begin{eqnarray}
&\displaystyle v_{\alpha x}=v_{\bot}\cos \phi,\qquad v_{\alpha
y}=\sqrt{\mu_{\alpha}} v_{\bot}\sin \phi, \qquad \phi=
\phi_{1}-\sqrt{\mu_{\alpha}}\omega_{c\alpha}t, \qquad
v_{z}=v_{\alpha z}\label{13}
\end{eqnarray}
where  $\mu_{\alpha}=1+V'_{0}/\omega_{c\alpha}>0$, the equation for
the perturbation $f_{\alpha}$ of the equilibrium distribution
$F_{0\alpha}$ function $(F_{\alpha}=F_{0\alpha}+f_{\alpha})$ has a
simple form,
\begin{eqnarray}
&\displaystyle \frac{\partial f_{\alpha}}{\partial
t}+\frac{e_{\alpha}}{m_{\alpha}\sqrt{\mu_{\alpha}}\omega_{c\alpha}}
\left(\frac{\partial\varphi}{\partial X_{\alpha}} \frac{\partial
f_{\alpha}} {\partial Y_{\alpha}}-\frac{\partial\varphi}{\partial
Y_{\alpha}} \frac{\partial f_{\alpha}} {\partial X_{\alpha}}\right)
+\frac{e_{\alpha}}{m_{\alpha}}\frac{\sqrt{\mu_{\alpha}}\omega_{c\alpha}}{v_{\perp}}
\left(\frac{\partial\varphi}{\partial \phi_{1}} \frac{\partial
f_{\alpha}} {\partial v_{\perp}}-\frac{\partial\varphi}{\partial
v_{\perp}}\frac{\partial
f_{\alpha}} {\partial \phi_{1}}\right)\nonumber \\
&\displaystyle
-\frac{e_{\alpha}}{m_{\alpha}}\frac{\partial\varphi}{\partial
z_{\alpha}} \frac{\partial f_{\alpha}}{\partial v_{z\alpha}}
=\frac{e_{\alpha}}{m_{\alpha}}\left[\frac{1} {\sqrt{\mu_{\alpha}}
\omega_{c\alpha}}\frac{\partial\varphi}{\partial Y_{\alpha}}
\frac{\partial F_{0\alpha}} {\partial X_{\alpha}}-
\frac{\sqrt{\mu_{\alpha}}\omega_{c\alpha}}{v_{\bot}}\frac{\partial\varphi}{\partial\phi_{\alpha}}
\frac{\partial F_{0\alpha}}{\partial
v_{\bot\alpha}}+\frac{\partial\varphi}{\partial z_{\alpha}}
\frac{\partial F_{0\alpha}}{\partial v_{z\alpha}}\right]. \label{14}
\end{eqnarray}
We assume that $|V'_{0}/\omega_{c\alpha}|\ll 1$ and put in what
follows $\mu_{\alpha}=1$. It is interesting to note, that the
equilibrium distribution function $F_{0\alpha}$, which in laboratory
frame contains the spatial inhomogeneity resulted from electric
field $\textbf{E}_{0}\left(\textbf{r}\right)$, does not contain such
inhomogeneity in convective coordinates (see Appendix 1). In what
follows we consider the equilibrium distribution function $F_{i0}$
as a Maxwellian,
\begin{eqnarray}
& \displaystyle F_{0\alpha}=\frac{n_{0\alpha}\left(X_{\alpha}
\right)}{\left(2\pi v^{2}_{T\alpha}\right)^{3/2}}
\exp\left(-\frac{v^{2}_{\bot}+v^{2}_{z}}{v^{2}_{T\alpha}} \right).
\label{15}
\end{eqnarray}
assuming the inhomogeneity of the density of plasma shear flow
species on coordinate $X_{\alpha}$. It follows from Eq.(\ref{8}), as
well as from Eq.(\ref{14}), that for $V'_{0}=const$, these equations
do not contain the spatial inhomogeneities, originated from
inhomogeneity of the flow velocity $V_{0}\left(x\right)$. Therefore,
the spatially homogeneous, but time dependent, Eqs.(\ref{8}) and
(\ref{14}) may be Fourier transformed over the variables
$x_{\alpha}, y_{\alpha}, z_{\alpha}$  with conjugate wave numbers
$k_{x}$, $k_{y}$ and $k_{z}$ and the temporal evolution of the
separate spatial Fourier mode of the perturbations of the
distribution function, $f_{\alpha}$ and of the electrostatic
potential, $\varphi\left(\textbf{k},t\right)$ may be traced upon. On
this way we present the potential $\varphi \left(\textbf{r}
,t\right) $ in form
\begin{eqnarray}
&\displaystyle \varphi \left(x_{\alpha}, y_{\alpha}, z_{\alpha},
t\right) = \int\varphi \left(k_{x}, k_{y}, k_{z} ,t
\right)e^{ik_{x}x_{\alpha}
+ik_{y}y_{\alpha}+ik_{z}z_{\alpha}} dk_{x}dk_{y}dk_{z}\nonumber  \\
&\displaystyle = \int\varphi \left(k_{x}, k_{y}, k_{z}, t \right)
\exp \left[ ik_{x}X_{\alpha}+ik_{y}Y_{\alpha}+ik_{z}z_{\alpha}\right.  \nonumber  \\
&\displaystyle \left. -i\frac{k_{\bot}\left(t\right)
v_{\bot}}{\omega_{c\alpha}}\sin\left( \phi-\omega_{c\alpha}t -\theta
\left(t\right)\right)\right]dk_{x}dk_{y}dk_{z},\label{16}
\end{eqnarray}
where
\begin{eqnarray}
&\displaystyle
k^{2}_{\perp}\left(t\right)=\left(k_{x}-V'_{0}tk_{y}\right)^{2}+k_{y}^{2}\label{17}
\end{eqnarray}
and $\tan \theta =k_{y}/(k_{x}-V'_{0}tk_{y})$. The solution to
linearized Eq.(\ref{14}) is calculated easily for any values of the
velocity shear rate $V'_{0}$ and it is equal to
\begin{eqnarray}
&\displaystyle f_{\alpha}\left(t,k_{x},k_{y}, k_{z},
v_{\bot},\phi,v_{z},z_{1}
\right)=\frac{ie_{\alpha}}{m_{\alpha}}\sum\limits_{n=-\infty}^{\infty}
\sum\limits_{n_{1}=-\infty}^{\infty} \int\limits^{t}dt_{1}
\varphi\left(t_{1},k_{x},k_{y},k_{z}\right) \nonumber  \\
&\displaystyle\times
\exp\Big(-ik_{z}v_{z}\left(t-t_{1}\right)+in\left(
\phi_{1}-\omega_{c}t-\theta\left(t\right)\right)-in_{1}\left(
\phi_{1}-\omega_{c}t_{1}-\theta\left(t\right)\right) \Big)\nonumber
\\ &\displaystyle\times
J_{n}\left(\frac{k_{\bot}\left(t\right)v_{\bot} }
{\omega_{c}}\right) J_{n_{1}}
\left(\frac{k_{\bot}\left(t_{1}\right)v_{\bot} } {\omega_{c}}\right)
\left[\frac{k_{y}}{\omega_{c\alpha}} \frac{\partial
F_{\alpha}}{\partial X_{\alpha}}+ \frac{\omega_{c}n_{1}}{v_{\bot}}
\frac{\partial F_{\alpha}}{\partial v_{\bot}}+ k_{1z}\frac{\partial
F_{\alpha}}{\partial v_{z}} \right]\nonumber  \\ &\displaystyle
+f_{\alpha}\left( t=t_{0},k_{x},k_{y}, k_{z}, v_{\perp}\phi, v_{z}
\right). \label{18}
\end{eqnarray}
The Poisson equation for separate spatial Fourier harmonic
$\varphi\left(\mathbf{k},t\right)$ gives the governing integral
equation\cite{Mikhailenko2010-1}
\begin{eqnarray}
&\displaystyle
\left[\left(k_{x}-V'_{0}tk_{y}\right)^{2}+k_{y}^{2}+k_{z}^{2}\right]
\varphi\left(\mathbf{k},t\right)=
\sum_{\alpha=i,e}\frac{i}{\lambda^{2}_{D\alpha}}\sum\limits_{n =
-\infty}^{\infty}\,\,\int\limits^{t}_{t_{0}}dt_{1}
\varphi\left(\mathbf{k},t_{1}\right)\nonumber  \\
&\displaystyle\times
I_{n}\left(k_{\perp}\left(t\right)k_{\perp}\left(t_{1}\right)\rho^{2}_{\alpha}\right)
e^{-\frac{1}{2}\rho^{2}_{\alpha}\left(k^{2}_{\perp}\left(t\right)+k^{2}_{\perp}
\left(t_{1}\right)\right)}
e^{-\frac{1}{2}k^{2}_{z}v^{2}_{T\alpha}\left(t-t_{1}\right)^{2}-in
\omega_{c\alpha}\left(t-t_{1}\right)
-in\left(\theta\left(t\right)-\theta\left(t_{1}\right)\right)}\nonumber \\
&\displaystyle \times\left[k_{y}v_{d\alpha}-n\omega_{c\alpha}
+ik^{2}_{z}v^{2}_{T\alpha}\left(t-t_{1}\right)\right]-4\pi\sum_{\alpha=i,e}e_{\alpha}\delta
n_{\alpha}\left(\mathbf{k}, t, t_{0}\right),\label{19}
\end{eqnarray}
where
\begin{eqnarray}
&\displaystyle 4\pi\sum_{\alpha=i,e}e_{\alpha}\delta
n_{\alpha}\left(\mathbf{k}, t, t_{0}\right)\nonumber \\
&\displaystyle=8\pi^{2}\sum_{\alpha=i,e}e_{\alpha}
\int\limits_{-\infty}^{\infty}dv_{z}e^{-ik_{z}v_{z}t}\int\limits_{0}^{\infty}
dv_{\bot}v_{\bot}J_{0}\left(\frac{k_{\perp}\left(t\right)v_{\bot}}{
\omega_{c\alpha}} \right)f_{\alpha}\left( t=t_{0}, \mathbf{k},
v_{\perp}, v_{z} \right),\label{20}
\end{eqnarray}
and where $f_{\alpha}\left( t=t_{0}, \mathbf{k}, v_{\perp}, v_{z}
\right)$ is the initial, determined at $t=t_{0}$ perturbation,
assumed here as not dependent on $\phi$, of the distribution
function $F_{\alpha}$. It follows from Eq.(\ref{19}), that initial
perturbation $\varphi\left(\mathbf{k},  t=t_{0}\right)$ of the
self-consistent electrostatic potential is equal to
\begin{eqnarray}
&\displaystyle \varphi\left(\mathbf{k},  t=t_{0}\right) =
-\frac{4\pi}{k^{2}_{\perp} \left(t_{0} \right)
+k^{2}_{z}}\sum_{\alpha=i,e}e_{\alpha}\delta
n_{\alpha}\left(\mathbf{k}, t_{0}, t_{0}\right).\label{21}
\end{eqnarray}
Eq.(\ref{19}) presents minimal model of the linear kinetic theory of
plasma shear flows, which incorporates the shear flow effects. In
Eq.(\ref{19}) these effects are concentrated in the time dependent
arguments of the Bessel function and manifest itself as a
time-dependent finite-Larmor-radius effects.

The integration by parts of the first term on the right of
Eq.(\ref{19}) gives the integral equation, which appears to be more
convenient and transparent for further analysis,
\begin{eqnarray}
&\displaystyle
\left[\left(k_{x}-V'_{0}tk_{y}\right)^{2}+k_{y}^{2}+k_{z}^{2}\right]\varphi
\left(\mathbf{k},t\right)=-\sum_{\alpha=i,e}\frac{1}{\lambda^{2}_{D\alpha}}
\varphi\left(\mathbf{k},t\right)+
\sum_{\alpha=i,e}\frac{1}{\lambda^{2}_{D\alpha}}\sum\limits_{n =
-\infty}^{\infty}\,\,\int\limits^{t}_{t_{0}}dt_{1}
\frac{d}{dt_{1}}\left\{\varphi\left(\mathbf{k},t_{1}\right)\right.\nonumber
\\ &\displaystyle\left.\times
I_{n}\left(k_{\perp}\left(t\right)k_{\perp}\left(t_{1}\right)
\rho^{2}_{\alpha}\right)
e^{-\frac{1}{2}\rho^{2}_{\alpha}\left(k^{2}_{\perp}\left(t\right)
+k^{2}_{\perp}\left(t_{1}\right)\right)
-in\left(\theta\left(t\right)-\theta\left(t_{1}\right)\right)}\right\}
e^{-\frac{1}{2}k^{2}_{z}v^{2}_{T\alpha}\left(t-t_{1}\right)^{2}
-in\omega_{c\alpha}\left(t-t_{1}\right)}\nonumber \\
&\displaystyle
+\sum_{\alpha=i,e}\frac{i}{\lambda^{2}_{D\alpha}}\sum\limits_{n =
-\infty}^{\infty}\,\,\int\limits^{t}_{t_{0}}dt_{1}
\varphi\left(\mathbf{k},t_{1}\right)k_{y}v_{d\alpha} \:
I_{n}\left(k_{\perp}\left(t\right)k_{\perp}\left(t_{1}\right)\rho^{2}_{\alpha}\right)\nonumber
\\ &\displaystyle\times
\exp\left[ -\frac{1}{2}\rho^{2}_{\alpha}\left(k^{2}_{\perp}\left(t\right)+k^{2}_{\perp}
\left(t_{1}\right)\right)
-in\left(\theta\left(t\right)-\theta\left(t_{1}\right)\right)-\frac{1}{2}
k^{2}_{z}v^{2}_{T\alpha}\left(t-t_{1}\right)^{2}-in\omega_{c\alpha}
\left(t-t_{1}\right)\right]  \nonumber \\
&\displaystyle -4\pi\sum_{\alpha=i,e}e_{\alpha}\delta
n_{\alpha}\left(\mathbf{k}, t,
t_{0}\right)+\sum_{\alpha=i,e}\frac{1}
{\lambda^{2}_{D\alpha}}\varphi\left(\mathbf{k},t_{0}\right)P_{\alpha}
\left(t, t_{0} \right),\label{22}
\end{eqnarray}
where $v_{d\alpha}=c T_{\alpha}/eBL_{n}$ is the diamagnetic drift
velocity, $L_{n}^{-1}=-d\ln n_{0\alpha}\left(x\right)/dx$,
$\rho_{\alpha}$ is thermal Larmor radius, and
\begin{eqnarray}
&\displaystyle P_{\alpha}\left(t, t_{0} \right)=\sum\limits_{n
=-\infty}^{\infty}I_{n}\left(k_{\perp}\left(t\right)k_{\perp}
\left(t_{0}\right)\rho^{2}_{\alpha}\right)\nonumber \\
&\displaystyle \times \exp\left[
-\frac{1}{2}\rho^{2}_{\alpha}\left(k^{2}_{\perp}
\left(t\right)+k^{2}_{\perp}\left(t_{0}\right)\right)
-in\left(\theta\left(t\right)-\theta\left(t_{0}\right)\right)
-\frac{1}{2}k^{2}_{z}v^{2}_{T\alpha}\left(t-t_{0}\right)^{2}
-in\omega_{c\alpha}\left(t-t_{0}\right)\right] .\label{23}
\end{eqnarray}
Note, that $P_{\alpha}\left(t_{0}, t_{0} \right)=1$.

It is important to note the alternative, explicitly causal
representation of Eq.(\ref{22}) with function $\Phi\left(\textbf{k},
t\right)=\varphi\left(\textbf{k},
t\right)\Theta\left(t-t_{0}\right)$, where
$\Theta\left(t-t_{0}\right)$ is the unit-step Heaviside function (it
is equal to zero for $t<t_{0}$ and equal to unity for $t\geq
t_{0}$). That equation for $\Phi\left(\textbf{k}, t\right)$ has a
form
\begin{eqnarray}
&\displaystyle
\left[\left(k_{x}-V'_{0}tk_{y}\right)^{2}+k_{y}^{2}+k_{z}^{2}\right]\Phi
\left(\mathbf{k},t\right)+4\pi\sum_{\alpha=i,e}e_{\alpha}\delta
n_{\alpha}\left(\mathbf{k}, t, t_{0}\right) \nonumber
\\ &\displaystyle =-\sum_{\alpha=i,e}\frac{1}{\lambda^{2}_{D\alpha}}
\int\limits^{t}_{t_{0}}dt_{1}
\frac{d}{dt_{1}}\Phi\left(\mathbf{k},t\right)+
\sum_{\alpha=i,e}\frac{1}{\lambda^{2}_{D\alpha}}\sum\limits_{n =
-\infty}^{\infty}\,\,\int\limits^{t}_{t_{0}}dt_{1}
\left.\frac{d}{dt_{1}}\right\{\Phi\left(\mathbf{k},t_{1}\right)\nonumber
\\ &\displaystyle\left.\times
I_{n}\left(k_{\perp}\left(t\right)k_{\perp}\left(t_{1}\right)
\rho^{2}_{\alpha}\right)
e^{-\frac{1}{2}\rho^{2}_{\alpha}\left(k^{2}_{\perp}\left(t\right)+k^{2}_{\perp}\left(t_{1}\right)\right)
-in\left(\theta\left(t\right)-\theta\left(t_{1}\right)\right)}\right\}
e^{-\frac{1}{2}k^{2}_{z}v^{2}_{T\alpha}\left(t-t_{1}\right)^{2}
-in\omega_{c\alpha}\left(t-t_{1}\right)}\nonumber \\
&\displaystyle
+\sum_{\alpha=i,e}\frac{i}{\lambda^{2}_{D\alpha}}\sum\limits_{n =
-\infty}^{\infty}\,\,\int\limits^{t}_{t_{0}}dt_{1}
\Phi\left(\mathbf{k},t_{1}\right)k_{y}v_{d\alpha}
I_{n}\left(k_{\perp}\left(t\right)k_{\perp}\left(t_{1}\right)\rho^{2}_{\alpha}\right)\nonumber
\\ &\displaystyle\times
\exp\left[
-\frac{1}{2}\rho^{2}_{\alpha}\left(k^{2}_{\perp}\left(t\right)+k^{2}_{\perp}
\left(t_{1}\right)\right)
-in\left(\theta\left(t\right)-\theta\left(t_{1}\right)\right)-\frac{1}{2}
k^{2}_{z}v^{2}_{T\alpha}\left(t-t_{1}\right)^{2}-in\omega_{c\alpha}
\left(t-t_{1}\right)\right] . \label{24}
\end{eqnarray}
Using the quasi neutrality approximation with
$\left(k^{2}_{\perp}\left(t\right)+k_{z}^{2}\right)\lambda^{2}_{D\alpha}\ll
1$, and averaging this equation over the time $t\gg
\omega_{ci}^{-1}$, we obtain from Eq.(\ref{24}) the equation, which
is relevant for the analysis of the low frequency drift type
perturbations,
\begin{eqnarray}
&\displaystyle \int\limits^{t}_{t_{0}}dt_{1}
\frac{d}{dt_{1}}\left\lbrace \Phi\left(\mathbf{k},t_{1}\right)
\left[ \left(1+\tau
\right)-I_{0}\left(k_{\perp}\left(t\right)k_{\perp}\left(t_{1}\right)\rho^{2}_{i}\right)
e^{-\frac{1}{2}\rho^{2}_{i}\left(k^{2}_{\perp}\left(t\right)+k^{2}_{\perp}
\left(t_{1}\right)\right)} \right]\right\rbrace \nonumber
\\ &\displaystyle
-i\int\limits^{t}_{t_{0}}dt_{1}
\Phi\left(\mathbf{k},t_{1}\right)k_{y}v_{di}
I_{0}\left(k_{\perp}\left(t\right)k_{\perp}\left(t_{1}\right)\rho^{2}_{i}\right)
e^{-\frac{1}{2}\rho^{2}_{i}\left(k^{2}_{\perp}\left(t\right)
+k^{2}_{\perp}\left(t_{1}\right)\right) } \nonumber
\\ &\displaystyle=
-\int\limits^{t}_{t_{0}}dt_{1} \frac{d}{dt_{1}}
\left(\Phi\left(\mathbf{k},t_{1}\right)
I_{0}\left(k_{\perp}\left(t\right)k_{\perp}\left(t_{1}\right)\rho^{2}_{i}\right)
e^{-\frac{1}{2}\rho^{2}_{i}\left(k^{2}_{\perp}\left(t\right)+k^{2}_{\perp}
\left(t_{1}\right)\right)}\right)
\left(1-e^{-\frac{1}{2}k^{2}_{z}v^{2}_{Ti}\left(t-t_{1}\right)^{2}}\right)\nonumber
\\ &\displaystyle
-i\int\limits^{t}_{t_{0}}dt_{1}
\Phi\left(\mathbf{k},t_{1}\right)k_{y}v_{di}
I_{0}\left(k_{\perp}\left(t\right)k_{\perp}\left(t_{1}\right)\rho^{2}_{i}\right)
e^{-\frac{1}{2}\rho^{2}_{i}\left(k^{2}_{\perp}\left(t\right)
+k^{2}_{\perp}\left(t_{1}\right)\right)}\left(1-e^{-\frac{1}{2}k^{2}_{z}v^{2}_{Ti}
\left(t-t_{1}\right)^{2}}\right) \nonumber
\\ &\displaystyle +\tau\int\limits^{t}_{t_{0}}dt_{1}
\left(\frac{d\Phi\left(\mathbf{k},t_{1}\right)}{dt_{1}}
+ik_{y}v_{de}\Phi\left(\mathbf{k},t_{1}\right)\right)e^{-\frac{1}{2}k^{2}_{z}v^{2}_{Te}
\left(t-t_{1}\right)^{2}},\label{25}
\end{eqnarray}
where $\tau=T_{i}/T_{e}$.  Eq.(\ref{25}) introduces obvious time
scales which determine different stages of the temporal evolution of
the electrostatic potential in plasma shear flows. As it follows
from Eq.(\ref{17}), the time dependence of $k_{\perp}\left(t\right)$
is negligible in times $t\ll \left(V'_{0}\right)^{-1}$ and may be
neglected; on that stage Eq.(\ref{25}) determines the ordinary modal
evolution of perturbations as in plasma without shear flow. In times
$\left(V'_{0}\right)^{-1}\ll t\ll t_{s}=
\left(V'_{0}k_{y}\rho_{i}\right)^{-1}$ time dependence of
$k_{\perp}$ gradually enhances the non-modal evolution of the
potential with time. In times $ t\gg t_{s}$, non-modality becomes
the dominant effect of the temporal evolution of the potential
$\Phi\left(\mathbf{k},t\right)$. In next Section we will obtain by
successive approximations the approximate solution to Eq.(\ref{25})
for long wavelength, $k_{\perp}\left(t_{0}\right)\rho_{i}<1$,
perturbations with weak ion Landau damping, for which
$\left|1-e^{-\frac{1}{2}k^{2}_{z}v^{2}_{Ti}\left(t-t_{0}\right)^{2}}\right|\ll
1$. The solution will be obtained without application the spectral
transformation over time for the case of a weak velocity shear, or a
small time, for which condition $|V'_{0}|\leq t\ll t_{s}$ is met,
and for long times, $ t\gg t_{s}$.

\section*{III.LINEAR NON-MODAL EVOLUTION OF THE \\ KINETIC  DRIFT INSTABILITY OF \\ PLASMA SHEAR FLOW}

If $k_{\perp}\left(t\right)\rho_{i}<1$ at time $t=t_{(0)}=0$ at
which the shear flow emerge , i.e.
$V'_{0}=V'_{0}\Theta\left(t\right)$, we will get
$k_{\perp}\left(t\right)\rho_{i}<1$ on times $t<t_{s}$ throughout.
By using the approximation
\begin{eqnarray}
&\displaystyle
I_{0}\left(k_{\perp}\left(t\right)k_{\perp}\left(t_{1}\right)\rho^{2}_{i}\right)
e^{-\frac{1}{2}\rho^{2}_{i}\left(k^{2}_{\perp}\left(t\right)
+k^{2}_{\perp}\left(t_{1}\right)\right)} \nonumber
\\ &\displaystyle\approx
b_{i}+\left(k_{x}k_{y}V'_{0}\rho^{2}_{i}\left(t+t_{1}\right)-\frac{1}{2}k^{2}_{y}\rho^{2}_{i}
\left(V'_{0}\right)^{2}\left(t^{2}+t_{1}^{2}\right)\right)\Theta\left(t\right)\label{26}
\end{eqnarray}
in Eq.(\ref{25}), we present Eq.(\ref{25}) in the form
\begin{eqnarray}
&\displaystyle \int\limits^{t}_{t_{0}}dt_{1} \left(\frac{d
\Phi\left(\mathbf{k},t_{1}\right)}{dt_{1}}+i\omega\left(\mathbf{k}\right)
\Phi\left(\mathbf{k},t_{1}\right)\right)\nonumber
\\ &\displaystyle=-\frac{b_{i}}{a_{i}}\int\limits^{t}_{t_{0}}dt_{1}\left(\frac{d
\Phi\left(\mathbf{k},t_{1}\right)}{dt_{1}}+ik_{y}v_{di}
\Phi\left(\mathbf{k},t_{1}\right)\right)\left(1-e^{-\frac{1}{2}k^{2}_{z}v^{2}_{Ti}
\left(t-t_{0}\right)^{2}}\right)\nonumber
\\ &\displaystyle+\frac{b_{i}}{a_{i}}\int\limits^{t}_{0}dt_{1}\left(\frac{d
\Phi\left(\mathbf{k},t_{1}\right)}{dt_{1}}+ik_{y}v_{di}
\Phi\left(\mathbf{k},t_{1}\right)\right)\left(\frac{k_{x}}{k_{y}}\frac{\left(t+t_{1}\right)}
{a_{i}V'_{0}t^{2}_{s}}-\frac{\left(t^{2}+t^{2}_{1}\right)}{2a_{i}t^{2}_{s}}\right)\nonumber \\
&\displaystyle+\int\limits^{t}_{0}dt_{1}\Phi\left(\mathbf{k},t_{1}\right)
\frac{1}{a_{i}V'_{0}t^{2}_{s}}\left(\frac{k_{x}}{k_{y}}-V'_{0}t\right)\nonumber \\
&\displaystyle-\frac{b_{i}}{a_{i}}\int\limits^{t}_{0}dt_{1}\left(\frac{d
\Phi\left(\mathbf{k},t_{1}\right)}{dt_{1}}+ik_{y}v_{di}
\Phi\left(\mathbf{k},t_{1}\right)\right)\left(\frac{k_{x}}{k_{y}}\frac{\left(t+t_{1}\right)}
{a_{i}V'_{0}t^{2}_{s}}-\frac{\left(t^{2}+t^{2}_{1}\right)}{2a_{i}t^{2}_{s}}\right)
\left(1-e^{-\frac{1}{2}k^{2}_{z}v^{2}_{Ti}
\left(t-t_{0}\right)^{2}}\right)\nonumber \\
&\displaystyle+\frac{\tau}{a_{i}}\int\limits^{t}_{t_{0}}dt_{1}\left(\frac{d
\Phi\left(\mathbf{k},t_{1}\right)}{dt_{1}}+ik_{y}v_{de}
\Phi\left(\mathbf{k},t_{1}\right)\right)\left(1-e^{-\frac{1}{2}k^{2}_{z}v^{2}_{Te}
\left(t-t_{0}\right)^{2}}\right),\label{27}
\end{eqnarray}
where $b_{i}=1-k^{2}_{\perp}\rho^{2}_{i}$,
$a_{i}=\tau+k^{2}_{\perp}\rho^{2}_{i}$ and
\begin{eqnarray}
&\displaystyle
\omega\left(\mathbf{k}\right)=-\frac{b_{i}}{a_{i}}k_{y}v_{di}.
\label{28}
\end{eqnarray}
The first term in the right side of Eq.(\ref{27}) determines the ion
Landau damping; this term is the same as in plasma without shear
flow. The next three terms originated from shear flow and determine
the corrections to the frequency and ion Landau damping provided by
shear flow. The right side of Eq.(\ref{27}) is small for
$\left(V'\right)^{-1}<t<t_{s}$, $\tau<1$ and for weak ion Landau
damping. Therefore the solution to Eq.(\ref{27}) we seek in the form
\begin{eqnarray}
&\displaystyle
\Phi\left(\mathbf{k},t\right)=C\exp\left(-i\omega\left(\mathbf{k}\right)t+i\nu
\left(\mathbf{k}, t\right)\right).\label{29}
\end{eqnarray}
Inserting Eq.(\ref{29}) into Eq.(\ref{27}) and neglecting the
derivative $d\nu \left(\mathbf{k}, t\right)/dt$ in the right side of
Eq.(\ref{27}), we obtain for $\nu \left(\mathbf{k}, t\right)$
following equation, assuming that $t_{0}\rightarrow -\infty$,
\begin{eqnarray}
&\displaystyle \int\limits^{t}_{t_{0}\rightarrow
-\infty}dt_{1}\Phi\left(\mathbf{k},t_{1}\right)\left[i\frac{d\nu
\left(\mathbf{k},
t_{1}\right)}{dt_{1}}-i\delta\omega\left(\mathbf{k}\right)-\frac{1}{a_{i}t^{2}_{s}}
\left(i\omega\left(\mathbf{k}\right)t_{1}^{2}\left(1+\frac{a_{i}}{b_{i}}\right)-t_{1}\right)\right]=0,
\label{30}
\end{eqnarray}
where
\begin{eqnarray}
&\displaystyle \delta\omega\left(\mathbf{k}\right)=
i\frac{\omega\left(\mathbf{k}\right)\left( \omega\left(\mathbf{k}
\right)
-k_{y}v_{di}\right)}{k_{z}v_{Ti}}\frac{b_{i}}{a_{i}}\sqrt{\frac{\pi}{2}}W\left(\frac{\omega\left(\mathbf{k}
\right)}{\sqrt{2}k_{z}v_{Ti}} \right)\nonumber \\ &\displaystyle
+i\tau \frac{\omega\left(\mathbf{k}\right)\left(
\omega\left(\mathbf{k} \right)
-k_{y}v_{de}\right)}{a_{i}k_{z}v_{Te}}\sqrt{\frac{\pi}{2}}W\left(\frac{\omega\left(\mathbf{k}
\right)}{\sqrt{2}k_{z}v_{Te}} \right) , \label{31}
\end{eqnarray}
with $\omega\left(\mathbf{k} \right)$ defined by Eq.(\ref{28}) and
$W\left(z\right)=\exp(-z^{2})\left(1
+\frac{2i}{\sqrt{\pi}}\int\limits^{z}_{0}d\zeta
e^{\zeta^{2}}\right)$. It is interesting to note, that $\text{Im}
\delta\omega\left(\mathbf{k}\right)=\gamma\left(\mathbf{k}\right)$
is right well known growth rate of the kinetic drift instability
\cite{Kadomtsev} determined in plasma theory approximately as
$\gamma\left(\mathbf{k}\right)=-\text{Im}\varepsilon\left(\mathbf{k},
\omega\left(\mathbf{k}\right)\right)/\left(\partial
\text{Re}\varepsilon\left(\mathbf{k},
\omega\left(\mathbf{k}\right)\right)/\partial\omega\left(\mathbf{k}\right)\right)$,
where $\varepsilon\left(\mathbf{k}, \omega\right)$ is ordinary
electrostatic dielectric permittivity of the plasma in magnetic
field. By equation to zero the expression in square brackets in
Eq.(\ref{30}), we obtain simple solution for $\nu \left(\mathbf{k},
t\right)$, which determines the non-modal evolution of the potential
$\Phi$ with time,
\begin{eqnarray}
&\displaystyle \Phi\left(\mathbf{k},t\right)=\Phi_{0}\exp
\left[-i\omega\left(\mathbf{k}\right)t\left(1-\frac{1+\tau}{a_{i}b_{i}}
\frac{t^{2}}{3t_{s}^{2}}\right)+iRe
\delta\omega\left(\mathbf{k}\right)t+\left(\gamma\left(\mathbf{k}\right)
-\frac{t}{2a_{i}t_{s}^{2}}\right)t\right].\label{32}
\end{eqnarray}
As it follows from Eq.(\ref{32}), non-modal effects, which reveal in
non-modal reduction of the frequency and growth rate, are negligible
at $t\ll t_{s}$ and become dominant at $t\sim t_{s}$. Note, that for
$\tau \gg k^{2}_{\perp}\rho^{2}_{i}$ the time $a^{1/2}_{i}t_{s}$ is
approximately equal to time $t_{2}=
\left(V'_{0}k_{\perp}\rho_{s}\right)^{-1} $ of the transition to
strongly non-modal regime in the fluid theory of the drift
turbulence of the plasma shear flow\cite{Mikhailenko2010}.

In the laboratory frame of reference  spatial Fourier mode
(\ref{32}) is observed as sheared mode with time dependent component
of the wave number $k_{x(lab)}= k_{x}-k_{y}V'_{0}t$ directed along
the velocity shear, and therefore is quite different from the normal
mode assumption,
\begin{eqnarray}
&\displaystyle \Phi\left(\mathbf{r},t\right)=\int dk_{x}\int
dk_{y}\int
dk_{z}\Phi_{0}\exp\left(ik_{x}x+ik_{y}y+ik_{z}z-ik_{y}V'_{0}tx\right)
\nonumber
\\ &\displaystyle \times\exp\left[-i\omega\left(\mathbf{k}\right)t\left(1-\frac{1+\tau}{a_{i}b_{i}}
\frac{t^{2}}{3t_{s}^{2}}\right)+iRe
\delta\omega\left(\mathbf{k}\right)t+\left(\gamma\left(\mathbf{k}\right)
-\frac{t}{2a_{i}t_{s}^{2}}\right)t\right].\label{33}
\end{eqnarray}
The mode shearing, which is other non-modal effect, becomes
pronounced in times $t\geq \gamma^{-1}$ when $V'_{0}\simeq \gamma$.
Only in times, for which $\left|V'_{0}t\right|\ll 1$ solution
(\ref{33}) has a normal mode form,
$\Phi\left(\mathbf{k},t\right)\sim \exp
\left(ik_{x}x+ik_{y}y+ik_{z}z-i\omega\left(\mathbf{k}\right)t\right)$.
Because of the time dependence $k_{x(lab)}=k_{x}-k_{y}V'_{0}t$ of
the wave number component along the flow shear, the modes in the
laboratory frame in times $t\gg \left(V'_{0}\right)^{-1}$ become
increasingly one--dimensional zonal--like as the perturbed $E\times
B$ velocity tilts more and more closely parallel to y-axis.

The exceptional advantage of the non-modal approach is a possibility
to perform the analysis of a plasma evolution on any finite time
domain. Now we consider the temporal evolution of drift
perturbations at times $t>t_{0}\gg t_{s}$ on the base of
Eq.(\ref{22}) averaged over the time $t\gg \omega^{-1}_{ci}$. Note,
that for the averaged Eq.(\ref{22}) ,
\begin{eqnarray}
&\displaystyle P_{i}\left(t, t_{0}
\right)=I_{0}\left(k_{\perp}\left(t\right)k_{\perp}
\left(t_{0}\right)\rho^{2}_{i}\right) \exp\left[
-\frac{1}{2}\rho^{2}_{i}\left(k^{2}_{\perp}
\left(t\right)+k^{2}_{\perp}\left(t_{0}\right)\right)
-\frac{1}{2}k^{2}_{z}v^{2}_{Ti}\left(t-t_{0}\right)^{2} \right]
,\label{34}
\end{eqnarray}
and $P_{i}\left(t_{0}, t_{0}
\right)=I_{0}\left(k^{2}_{\perp}\left(t_{0}\right)\rho^{2}_{i}\right)
e^{-\rho^{2}_{i}k^{2}_{\perp} \left(t_{0}\right)}$. We consider the
times $t>t_{0}\gg t_{s}$, for which $k_{\perp}\left(t \right)$
becomes large enough, so that $k_{\perp}\left(t
\right)\rho_{i}>k_{\perp}\left(t_{0}\right)\rho_{i}>1$ and
$P_{i}\left(t_{0}, t_{0}\right)\simeq
\left(\sqrt{2\pi}k_{\perp}\left(t_{0}\right)\rho_{i}\right)^{-1}$.
The initial value $\varphi\left(\mathbf{k},t=t_{0}\right)$ for the
averaged potential in that case have to be corrected and is equal to
\begin{eqnarray}
&\displaystyle\varphi\left(\mathbf{k},t=t_{0}\right)
=\varphi\left(\mathbf{k},t_{0}\right)\frac{t_{s}}{\sqrt{2\pi}
t_{0}},\label{35}
\end{eqnarray}
where $\varphi\left(\mathbf{k},t_{0}\right)$ is the initial value
for Eq.(\ref{22}) without averaging. For the times considered
\begin{eqnarray}
&\displaystyle I_{0}\left(k_{\perp}\left(t\right)k_{\perp}
\left(t_{1}\right)\rho^{2}_{i}\right)
e^{-\frac{1}{2}\rho^{2}_{i}\left(k^{2}_{\perp}
\left(t\right)+k^{2}_{\perp}\left(t_{1}\right)\right)
-\frac{1}{2}k^{2}_{z}v^{2}_{Ti}\left(t-t_{1}\right)^{2}}\approx
\frac{t_{s}}{\sqrt{2\pi
tt_{1}}}\exp\left(-\frac{1}{2}\kappa_{i}^{2}\left(t-t_{1}\right)^{2}\right),
\label{36}
\end{eqnarray}
where
\begin{eqnarray}
&\displaystyle
\kappa_{i}^{2}=\frac{1}{t^{2}_{s}}+k^{2}_{z}v^{2}_{Ti}. \label{37}
\end{eqnarray}
With this approximation, Eq.(\ref{22}) becomes
\begin{eqnarray}
&\displaystyle \left(1+\tau \right)\int\limits^{t}_{t_{0}}dt_{1}
\frac{d\varphi\left(\mathbf{k},t_{1}\right)}{dt_{1}}=\int\limits^{t}_{t_{0}}dt_{1}
\frac{d}{dt_{1}}\left[\varphi\left(\mathbf{k},t_{1}\right)\frac{t_{s}}{\left(2\pi
tt_{1}\right)^{1/2}}e^{-\frac{1}{2t^{2}_{s}}\left(t-t_{1}\right)^{2}}\right]
e^{-\frac{1}{2}k_{z}^{2}v^{2}_{Ti}\left(t-t_{1}\right)^{2}}\nonumber
\\ &\displaystyle +ik_{y}v_{di}\int\limits^{t}_{t_{0}}dt_{1}\varphi\left(\mathbf{k},t_{1}\right)
\frac{t_{s}}{\left(2\pi
tt_{1}\right)^{1/2}}e^{-\frac{\kappa^{2}_{i}}{2}\left(t-t_{1}\right)^{2}}\nonumber
\\ &\displaystyle+\tau\int\limits^{t}_{t_{0}}dt_{1}\left(
\frac{d\varphi\left(\mathbf{k},t_{1}\right)}{dt_{1}}+ik_{y}v_{de}
\varphi\left(\mathbf{k},t_{1}\right)\right)e^{-\frac{1}{2}k_{z}^{2}v^{2}_{Te}\left(t-t_{1}\right)^{2}}
\nonumber \\ &\displaystyle
+\varphi\left(\mathbf{k},t_{0}\right)\frac{t_{s}}{\left(2\pi
tt_{0}\right)^{1/2}}\left(e^{-\frac{\kappa^{2}_{i}}{2}\left(t-t_{0}\right)^{2}}
-\left(\frac{t}{t_{0}}\right)^{1/2}\right). \label{38}
\end{eqnarray}
In Eq.(\ref{38}), electron term may be omitted, because for times
$t-t_{0}>t_{s}$ and  $V'_{0}\sim \gamma=
k_{y}v_{di}k_{\perp}^{2}\rho_{i}^{2}$, where $\gamma$ is the growth
rate of the drift kinetic instability,
\begin{eqnarray}
&\displaystyle
\exp\left(-\frac{1}{2}k_{z}^{2}v^{2}_{Te}\left(t-t_{1}\right)^{2}\right)
<\exp\left(-\frac{1}{2}k_{z}^{2}v^{2}_{Te}t^{2}_{s}\right)\sim\exp\left(-\frac{k^{2}_{z}}{\tau
k^{2}_{y}}\frac{m_{i}}{m_{e}}\frac{k^{2}_{\perp}L_{n}^{2}}{\left(k_{\perp}\rho_{i}\right)^{4}}\right)\ll
1.\label{39}
\end{eqnarray}
In zero approximation over $t_{s}/t$ we have for
$\varphi\left(\mathbf{k},t\right)$ equation
$\int\limits^{t}_{t_{0}}dt_{1}
d\varphi\left(\mathbf{k},t_{1}\right)/dt_{1}=0$ with solution
$\varphi\left(\mathbf{k},t_{1}\right)/dt_{1}=\varphi_{0}=const$.
Accounting for in first approximation small right hand side of
Eq.(\ref{38}), we seek solution for
$\varphi\left(\mathbf{k},t\right)$ in the form
\begin{eqnarray}
&\displaystyle
\varphi\left(\mathbf{k},t\right)=\varphi_{0}\exp\left(\nu\left(\mathbf{k},t\right)\right),\label{40}
\end{eqnarray}
where $\nu\left(\mathbf{k},t\right)= O\left(t_{s}/t\right)$. For
times $t\geq t_{0}\gg \kappa^{-1}$ we can omit in Eq.(\ref{38}) the
terms with small exponents
$\exp\left(-(\kappa^{2}_{i}/2)\left(t-t_{0}\right)^{2}\right)$ and
obtain the following solution for
$\varphi\left(\mathbf{k},t\right)$:
\begin{eqnarray}
&\displaystyle
\varphi\left(\mathbf{k},t\right)=\varphi_{0}\exp\left(\frac{1}{\sqrt{2\pi}\kappa_{i}^{2}t_{s}t}
+i\frac{k_{y}v_{di}t_{s}}{2\kappa_{i}t}\right),\label{41}
\end{eqnarray}
which gradually becomes a zero-frequency cell-like perturbation.
\section*{IV.NONLINEAR PHASE SHIFT IN SHEAR FLOW}
Nonlinear terms in the left hand side of Eq.(\ref{14}) can
drastically change linear non-modal solutions (\ref{33}) and
(\ref{41}). Because of the nonlinearities, accounted in the left
side of Eq.(\ref{14}), variables $X, Y, v_{\perp}, \phi$ and $z$
become coupled and determined by equations of characteristics,
\begin{eqnarray}
& \displaystyle
dt=\frac{dX}{-\dfrac{e}{m_{i}\omega_{ci}}\dfrac{\partial\varphi}{\partial
Y}}=
\frac{dY}{\dfrac{e}{m_{i}\omega_{ci}}\dfrac{\partial\varphi}{\partial
X}}= \frac{dv_{\bot}}{\dfrac{e}{m_{i}}\dfrac{\omega_{ci}}{v_{\bot}}
\dfrac{\partial\varphi}{\partial
\phi_{1}}}=\frac{d\phi_{1}}{-\dfrac{e}{m_{i}}\dfrac{\omega_{ci}}{v_{\bot}}
\dfrac{\partial\varphi}{\partial
v_{\bot}}}=\frac{dv_{z}}{-\dfrac{e}{m_{i}}\dfrac{\partial\varphi}{\partial
z_{1}}} \nonumber\\ & \displaystyle
=\frac{df_{i}}{\dfrac{e}{m_{i}\omega_{ci}}\dfrac{\partial\varphi}{\partial
Y} \dfrac{\partial F _{i0}}{\partial
X}-\dfrac{e}{m_{i}}\dfrac{\omega_{ci}}{v_{\bot}}
\dfrac{\partial\varphi}{\partial \phi_{1}} \dfrac{\partial F
_{i0}}{\partial v_{\bot}}
+\dfrac{e}{m_{i}}\dfrac{\partial\varphi}{\partial z_{1}}
\dfrac{\partial F _{i0}}{\partial v_{z}}}.\label{42}
\end{eqnarray}
Last equation in system (\ref{42}) gives nonlinear solution for the
perturbation of the ion distribution function $f_{i}$ with known
$F_{i0}$,
\begin{eqnarray}
& \displaystyle f_{i} =
\frac{e}{m}\int\limits^{t}\left[\frac{1}{\omega_{ci}}\frac{\partial\varphi}{
\partial Y}\frac{\partial F_{i0}}{\partial X}-\frac{\omega_{ci}}{v_{\bot}}
\frac{\partial\varphi}{\partial \phi_{1}} \frac{\partial
F_{i0}}{\partial v_{\bot}} +\frac{\partial\varphi}{\partial z_{1}}
\frac{\partial F _{i0}}{\partial v_{z}} \right] dt', \label{43}
\end{eqnarray}
in which coordinates $X$, $Y$, $v_{\perp}$, $\phi$  are
$X=\bar{X}+\delta X$, $Y=\bar{Y}+\delta Y$,
$\phi=\bar{\phi}+\delta\phi$, where $\bar{X}$, $\bar{Y}$, are the
guiding center coordinates averaged over the turbulent pulsations,
and $\delta X(t)$, $\delta Y(t)$, $\delta\phi$ are random ion orbit
disturbances due to their scattering by electrostatic low frequency
drift turbulence. The disturbances  are assumed sufficiently small
and, after the averaging over the times $t\gg
\left(\omega_{ci}\right)^{-1}$, they are determined by the equations
\begin{eqnarray}
& \displaystyle \delta
X=-\dfrac{e}{m_{i}\omega_{ci}}\int\limits^{t}_{t_{0}}\dfrac{\partial\varphi}
{\partial\bar{Y}}dt_{1}
=-\frac{c}{B}\int\limits^{t}_{t_{0}}dt_{1}\int
d\mathbf{k}\varphi\left(\mathbf{k},t_{1}\right)k_{y}J_{0}\left(\frac{k_{\perp}
\left(t_{1}\right)v_{\perp}}{\omega_{ci}}\right)e^{i\Psi},\label{44}
\\
& \displaystyle\delta
Y=\dfrac{e}{m_{i}\omega_{ci}}\int\limits^{t}_{t_{0}}\dfrac{\partial\varphi}
{\partial \bar{X}}dt_{1}
=\frac{c}{B}\int\limits^{t}_{t_{0}}dt_{1}\int
d\mathbf{k}\varphi\left(\mathbf{k},t_{1}\right)k_{x}J_{0}\left(\frac{k_{\perp}
\left(t_{1}\right)v_{\perp}}{\omega_{ci}}\right)e^{i\Psi},\label{45}
\\
&\displaystyle\delta\phi=-\dfrac{e}{m_{i}}\dfrac{\omega_{ci}}{v_{\bot}}
\int\limits^{t}_{t_{0}}\dfrac{\partial\varphi}{\partial
\bar{v}_{\bot}}dt_{1}
=\frac{e}{mv_{\perp}}\int\limits^{t}_{t_{0}}dt_{1}\int
d\mathbf{k}\varphi\left(\mathbf{k},t_{1}\right)k_{\perp}
\left(t_{1}\right)J_{1}\left(\frac{k_{\perp}
\left(t_{1}\right)v_{\perp}}{\omega_{ci}}\right)e^{i\Psi},\label{46}
\end{eqnarray}
and $\delta v_{\bot}=0$. In Eqs.(\ref{44})--(\ref{46}),
$\Psi=k_{x}X+k_{y}Y+k_{z}z
+\mathbf{k}\left(t_{1}\right)\delta\mathbf{r}\left(t_{1}\right)$,
and $i\mathbf{k}\left(t\right)\delta\mathbf{r}\left(t\right)$
denotes the phase shift resulted from perturbations of the ions
orbits due to random ion-waves interactions,
\begin{eqnarray}
& \displaystyle
\mathbf{k}\left(t\right)\delta\mathbf{r}\left(t\right)=k_{x}\delta
X\left(t\right)+k_{y}\delta Y\left(t\right)
-\frac{k_{\bot}\left(t\right)\bar{v}_{\bot}}{\omega_{ci}}\cos
\left(\phi-\theta\right)\delta\phi\left(t\right). \label{47}
\end{eqnarray}
In Eq.(\ref{47}), the scattering of ions along the magnetic field is
ignored. In variables $\bar{X}$, $\bar{Y}$, $\bar{\phi}$,
$v_{\bot}$, the averaged over the time $t\gg \omega^{-1}_{ci}$ and
over initial phases of the drift perturbations solution for
$f_{\alpha}\left(t,k_{x},k_{y}, k_{z}, v_{\bot},\phi,v_{z},z_{1}
\right)$ in  drift frequency range has a form
\begin{eqnarray}
&\displaystyle f_{\alpha}\left(t,k_{x},k_{y}, k_{z}, v_{\bot
},\phi,v_{z},z_{1} \right)=i\frac{e_{\alpha}}{m_{\alpha}}
\int\limits^{t}dt_{1}
\varphi\left(t_{1},k_{x},k_{y},k_{z}\right) \nonumber  \\
&\displaystyle\times
\exp\Big(-ik_{z}v_{z}\left(t-t_{1}\right)-\frac{1}{2}\left\langle\left(\mathbf{k}\left(t\right)
\delta\mathbf{r}\left(t\right)-\mathbf{k}\left(t_{1}\right)
\delta\mathbf{r}\left(t_{1}\right)\right)^{2}\right\rangle\Big)\nonumber \\
&\displaystyle\times J_{0}\left(\frac{k_{\bot}\left(t\right)v_{\bot}
} {\omega_{c}}\right) J_{0}
\left(\frac{k_{\bot}\left(t_{1}\right)v_{\bot} } {\omega_{c}}\right)
\left[\frac{k_{y}}{\omega_{c\alpha}} \frac{\partial
F_{\alpha}}{\partial X_{\alpha}}+ k_{1z}\frac{\partial
F_{\alpha}}{\partial v_{z}} \right]\nonumber
\\ &\displaystyle +f_{\alpha}\left( t=t_{0},k_{x},k_{y}, k_{z},
v_{\perp}\phi, v_{z} \right), \label{48}
\end{eqnarray}
in which $f_{\alpha}\left( t=t_{0}, \mathbf{k}, v_{\perp}, v_{z}
\right)$ is the initial, determined at $t=t_{0}$ perturbation,
assumed as independent on $\phi$, of the distribution function
$F_{\alpha}$. In Eq.(\ref{48}) Gaussian distribution for ions orbit
disturbances is assumed. We use Eq.(\ref{48}) in Poisson equation
for the potential $\varphi\left(\textbf{r}_{\alpha},t\right)$,
\begin{eqnarray}
&\displaystyle \vartriangle \varphi\left(\textbf{r},t\right)=
-4\pi\sum_{\alpha=i,e} e_{\alpha}\int f_{\alpha}\left(\textbf{v},
\textbf{r}, t \right)d\textbf {v}_{\alpha}, \label{49}
\end{eqnarray}
and obtain integral equation for separate spatial Fourier harmonic
$\varphi\left(\mathbf{k},t\right)$ for electrostatic potential, in
which effect of the turbulent scattering of ions on sheared drift
modes of random phases is accounted for,
\begin{eqnarray}
&\displaystyle k^{2}\left(t\right)\varphi\left(\mathbf{k},t\right)=
\sum_{\alpha=i,e}\frac{i}{\lambda^{2}_{D\alpha}v^{2}_{T\alpha}}\int\limits^{t}_{t_{0}}dt_{1}
\varphi\left(\mathbf{k},t_{1}\right)\int\limits_{0}^{\infty}
dv_{\bot}v_{\bot}\exp\left( -\frac{v^{2}_{\bot}}{v^{2}_{T\alpha}}\right) \nonumber  \\
&\displaystyle\times J_{0}\left(\frac{k_{\bot}\left(t\right)v_{\bot}
} {\omega_{c}}\right) J_{0}
\left(\frac{k_{\bot}\left(t_{1}\right)v_{\bot} }
{\omega_{c}}\right)\exp\left(-\frac{1}{2}k^{2}_{z}v^{2}_{T\alpha}\left(t-t_{1}\right)^{2}
-\frac{1}{2}\left\langle\left(\mathbf{k}\left(t\right)
\delta\mathbf{r}\left(t\right)-\mathbf{k}\left(t_{1}\right)
\delta\mathbf{r}\left(t_{1}\right)\right)^{2}\right\rangle\right)\nonumber \\
&\displaystyle \times\left[k_{y}v_{d\alpha}
+ik^{2}_{z}v^{2}_{T\alpha}\left(t-t_{1}\right)\right]-4\pi\sum_{\alpha=i,e}e_{\alpha}\delta
n_{\alpha}\left(\mathbf{k}, t, t_{0}\right).\label{50}
\end{eqnarray}

It follows from Eqs.(\ref{44})--(\ref{46}), that for $k_{x}\sim
k_{y}$ we have the estimates $\delta X\sim \delta Y$ and
\begin{eqnarray}
&\displaystyle \left|\frac{k_{x}\delta X\left(t\right)}{k_{\perp}
\left(t\right)\dfrac{v_{\perp}}{\omega_{c}}\delta\phi\left(t\right)}\right|\sim
\left|\frac{k^{2}_{y}J_{0}\left(\dfrac{k_{\perp}
\left(t_{1}\right)v_{\perp}}{\omega_{ci}}\right)}{k^{2}_{\perp}
\left(t_{1}\right) J_{1}\left(\dfrac{k_{\perp}
\left(t_{1}\right)v_{\perp}}{\omega_{ci}}\right)}\right|.\label{51}
\end{eqnarray}
At initial times of the evolution, $t\ll \left(V_{0}'\right)^{-1}$,
for long wavelength perturbations with $k_{\perp}
\left(t\right)v_{\perp}\ll \omega_{ci}$, the nonlinear phase shift
is determined mainly by $\delta X$ and $\delta Y$. For short
wavelength perturbations with $k_{\perp} \left(t\right)v_{\perp}>
\omega_{ci}$ for these times the $\delta\phi\left(t\right)$ term is
important as well. At these times non-modal effects are negligible.
At times $t> \left(V_{0}'\right)^{-1}$ for $k_{\perp}
\left(t\right)v_{\perp}\ll \omega_{ci}$
\begin{eqnarray}
&\displaystyle \left|\frac{k_{x}\delta X\left(t\right)}{k_{\perp}
\left(t\right)\dfrac{v_{\perp}}{\omega_{c}}\delta\phi\left(t\right)}
\right|\sim\frac{1}{k_{y}\rho_{i}}\frac{1}{\left(V'_{0}t\right)^{3}},\label{52}
\end{eqnarray}
and for $k_{\perp} \left(t\right)v_{\perp}> \omega_{ci}$,
\begin{eqnarray}
&\displaystyle \left|\frac{k_{x}\delta X\left(t\right)}{k_{\perp}
\left(t\right)\dfrac{v_{\perp}}{\omega_{c}}\delta\phi\left(t\right)}
\right|\sim\frac{1}{\left(V'_{0}t\right)^{2}}.\label{53}
\end{eqnarray}
It follows from Eqs.(\ref{52}), (\ref{53}), that at times $t>
\left(V_{0}'\right)^{-1}$ turbulent scattering of the angle
$\delta\phi\left(t\right)$ is the dominant process in the formation
the turbulent shift of the phase of the electrostatic potential. Now
consider the average phase shift term for times
$\left(V'_{0}k_{y}\rho_{i}\right)^{-1}> t>
\left(V'_{0}\right)^{-1}$, but for
$k_{\perp}v_{\perp}/\omega_{ci}<1$. In that case
$k_{\perp}\left(t\right)\approx k_{y}V'_{0}t$ and
\begin{eqnarray}
&\displaystyle
J_{1}\left(\frac{k_{\perp}\left(t\right)v_{\perp}}{\omega_{ci}}\right)\approx
\frac{k_{\perp}\left(t\right)v_{\perp}}{2\omega_{ci}}\approx
\frac{k_{y}V'_{0}tv_{\perp}}{\omega_{ci}}.\label{54}
\end{eqnarray}
In convective coordinates, solution for potential
$\varphi\left(\mathbf{k},t\right)$ at $k_{y}V'_{0}t\rho_{i}<1$ (and
$V'_{0}t>1$) has a modal form
\begin{eqnarray}
&\displaystyle
\varphi\left(\mathbf{k},t\right)=\varphi\left(\mathbf{k},t_{0}\right)
e^{i\omega\left(\mathbf{k}\right)t+\gamma\left(\mathbf{k}\right)t},\label{55}
\end{eqnarray}
where $\omega\left(\mathbf{k}\right)$ and
$\gamma\left(\mathbf{k}\right)$ are the frequency and growth rate of
the kinetic drift instability. For times
$t>\left(V'_{0}\right)^{-1}$, the approximation
\begin{eqnarray}
&\displaystyle\left\langle\left(\mathbf{k}\left(t\right)
\delta\mathbf{r}\left(t\right)-\mathbf{k}\left(t_{1}\right)
\delta\mathbf{r}\left(t_{1}\right)\right)^{2}\right\rangle
\approx\frac{v_{\bot}^{2}}{2\omega^{2}_{ci}}\left\langle\left(\mathbf{k}\left(t\right)
\delta\phi\left(t\right)-\mathbf{k}\left(t_{1}\right)
\delta\phi\left(t_{1}\right)\right)^{2}\right\rangle \nonumber\\ &
\displaystyle \approx\frac{v_{\bot}^{2}}{2\omega^{2}_{ci}}
\left\langle\left(\mathbf{k}\left(t\right)\left(\delta\phi\left(t\right)
-\delta\phi\left(t_{1}\right)\right)\right)^{2}\right\rangle\nonumber\\
& \displaystyle \approx
\frac{e^{2}k^{2}_{y}\left(V'_{0}\right)^{6}v^{2}_{\perp}t^{2}}{8\omega^{4}_{ci}m^{2}_{i}}
\int\limits^{t}_{t_{1}}dt'_{1}\int\limits^{t}_{t_{1}}dt'_{2}\int
d\mathbf{k}_{1}\left|\varphi\left(\mathbf{k}_{1},
t_{0}\right)\right|^{2}k^{4}_{1y}\exp\Big[\gamma
\left(\mathbf{k}_{1}\right)\left(t'_{1}+t'_{2}\right)+i\omega\left(\mathbf{k}_{1}\right)
\left(t'_{1}-t'_{2}\right)\Big] \nonumber\\
& \displaystyle
\times\left(t'_{1}t'_{2}\right)^{2}\exp\left[ -\frac{1}{2}\left\langle\mathbf{k}\left(t'_{1}\right)
\left(\delta\textbf{r}\left(t'_{1}\right)-\delta\textbf{r}\left(t'_{2}\right)\right)^{2}\right\rangle\right] \label{56}
\end{eqnarray}
is valid. With time variables $\tau=t_{1}-t_{2}$ and
$\widehat{t}=\left(t_{1}+t_{2}\right)/2$, Eq.(\ref{56}) becomes
\begin{eqnarray}
&\displaystyle \frac{v_{\bot}^{2}}{2\omega^{2}_{ci}}
\left\langle\left(\mathbf{k}\left(t\right)\left(\delta\phi\left(t\right)
-\delta\phi\left(t_{1}\right)\right)\right)^{2}\right\rangle=\frac{c^{2}k^{2}_{y}
\left(V'_{0}\right)^{6}v_{\bot}^{2}t^{2}}{8B^{2}\omega_{ci}^{2}}\left(
\int\limits^{0}_{t_{1}-t}d\tau\int\limits^{t+\frac{\tau}{2}}_{t_{1}-
\frac{\tau}{2}}d\hat{t}+\int\limits^{t-t_{1}}_{0}
d\tau\int\limits^{t-\frac{\tau}{2}}_{t_{1}+\frac{\tau}{2}}d\hat{t}\right)
\nonumber
\end{eqnarray}
\begin{eqnarray}
& \displaystyle \times\int
d\mathbf{k}_{1}\left|\varphi\left(\mathbf{k}_{1},
t_{0}\right)\right|^{2}k^{4}_{1y}\hat{t}^{4}e^{2\gamma
\left(\mathbf{k}_{1}\right)\hat{t}+i\omega\left(\mathbf{k}_{1}\right)\tau}
\exp\left[-\frac{c^{2}k^{2}_{1y}
\left(V'_{0}\right)^{6}v_{\bot}^{2}t^{2}}{16B^{2}\omega_{ci}^{2}}\right.\nonumber\\
& \displaystyle\left.\times\left(
\int\limits^{0}_{-\tau}d\tau_{1}\int\limits^{\hat{t}+\frac{1}{2}\left(\tau+\tau_{1}\right)}
_{\hat{t}-\frac{1}{2}\left(\tau+\tau_{1}\right)}d\hat{t}_{1}
+\int\limits^{\tau}_{0}d\tau_{1}\int\limits^{\hat{t}+\frac{1}{2}\left(\tau-\tau_{1}\right)}
_{\hat{t}-\frac{1}{2}\left(\tau-\tau_{1}\right)}d\hat{t}_{1}\right)
\int d\mathbf{k}_{2}\left|\varphi\left(\mathbf{k}_{2},
t_{0}\right)\right|^{2}k^{4}_{2y}\hat{t}^{4}_{1}
\right.\nonumber\\
& \displaystyle\left.\times \exp\left( 2\gamma
\left(\mathbf{k}_{2}\right)\hat{t}_{1}+i\omega\left(\mathbf{k}_{2}\right)
\tau_{1}\right.\right.\nonumber\\
& \displaystyle\left.\left.-\frac{1}{2}\left\langle\left(
\mathbf{k}\left(\hat{t_{1}}+\frac{1}{2}\tau_{1}\right)
\left(\delta\textbf{r}\left(\hat{t_{1}}+\frac{1}{2}\tau_{1}\right)-
\delta\textbf{r}\left(\hat{t_{1}}-\frac{1}{2}\tau_{1}\right)\right)\right)^{2}\right\rangle\right)
\right]. \label{57}
\end{eqnarray}
The integration over $\hat{t}_{1}$ is performed over narrow interval
$\left(\hat{t}+\frac{1}{2}\left(\tau\pm\tau_{1}\right),
\hat{t}-\frac{1}{2}\left(\tau\pm\tau_{1}\right)\right)$. In such a
case the integrals over $\hat{t_{1}}$ may be approximately
calculated as
\begin{eqnarray}
&\displaystyle\int\limits_{\hat{t}-\frac{1}{2}\left(\tau\pm\tau_{1}\right)}
^{\hat{t}+\frac{1}{2}\left(\tau\pm\tau_{1}\right)}
d\hat{t_{1}}\hat{t}_{1}^{4}e^{2\gamma\left(\mathbf{k}_{2}\right)\hat{t}_{1}-f\left(\hat{t}_{1},
\tau\right)}\approx
\left(\tau\pm\tau_{1}\right)\hat{t}^{4}e^{2\gamma
\left(\mathbf{k}_{2}\right)\hat{t}-f\left(\hat{t},
\tau\right)},\label{58}
\end{eqnarray}
and for Eq.(\ref{57}) we obtain
\begin{eqnarray}
&\displaystyle \frac{v_{\bot}^{2}}{2\omega^{2}_{ci}}
\left\langle\left(\mathbf{k}\left(t\right)\left(\delta\phi\left(t\right)
-\delta\phi\left(t_{1}\right)\right)\right)^{2}\right\rangle=\frac{c^{2}k^{2}_{y}
\left(V'_{0}\right)^{6}v_{\bot}^{2}t^{2}}{8B^{2}\omega_{ci}^{2}}\left[
\int\limits^{0}_{t_{1}-t}d\tau\int\limits^{t+\frac{\tau}{2}}_{t_{1}-
\frac{\tau}{2}}d\hat{t}+\int\limits^{t-t_{1}}_{0}
d\tau\int\limits^{t-\frac{\tau}{2}}_{t_{1}+\frac{\tau}{2}}d\hat{t}\right]
\nonumber\\& \displaystyle \int
d\mathbf{k}_{1}\left|\varphi\left(\mathbf{k}_{1},
t_{0}\right)\right|^{2}k^{4}_{1y}\hat{t}^{4}e^{2\gamma
\left(\mathbf{k}_{1}\right)\hat{t}+i\omega\left(\mathbf{k}_{1}\right)\tau}\nonumber\\&
\displaystyle
\exp\left(-\tau\frac{c^{2}k^{2}_{1y}\left(V'_{0}\right)^{6}v_{\bot}^{2}\hat{t}^{6}}
{8B^{2}\omega_{ci}^{2}}\int
d\mathbf{k}_{2}k^{4}_{2y}\left|\varphi\left(\mathbf{k}_{2},
t_{0}\right)\right|^{2}e^{2\gamma
\left(\mathbf{k}_{2}\right)\hat{t}}\frac{C\left(\mathbf{k}_{2},\hat{t}\right)}
{\omega^{2}\left(\mathbf{k}_{2}\right)}\right),\label{59}
\end{eqnarray}
where $C\left(\mathbf{k}_{2},\hat{t}\right)$ resulted from infinite
sequences of inner integration in exponential of the \newline
$\left\langle\mathbf{k}\left(\hat{t_{i}}+\frac{1}{2}\tau_{i}\right)
\left(\delta\textbf{r}\left(\hat{t_{i}}+\frac{1}{2}\tau_{i}\right)-
\delta\textbf{r}\left(\hat{t_{i}}-\frac{1}{2}\tau_{i}\right)\right)^{2}\right\rangle$
over $\hat{t}_{i}$ and determined by integral equation
\begin{eqnarray}
&\displaystyle
C\left(\mathbf{k}_{1},\hat{t}\right)=\frac{c^{2}k^{2}_{1y}\left(V'_{0}\hat{t}\right)^{6}v_{\bot}^{2}}
{8B^{2}\omega_{ci}^{2}}\int
d\mathbf{k}_{2}k^{4}_{2y}\left|\varphi\left(\mathbf{k}_{2},
t_{0}\right)\right|^{2}e^{2\gamma
\left(\mathbf{k}_{2}\right)\hat{t}}\frac{C\left(\mathbf{k}_{2},\hat{t}\right)}
{\omega^{2}\left(\mathbf{k}_{2}\right)}.\label{60}
\end{eqnarray}
Changing $\tau \rightarrow -\tau$ in first integral over $\tau$ in
Eq.(\ref{59}), we perform the integration over $\tau$ and obtain
simple result
\begin{eqnarray}
&\displaystyle \frac{v_{\bot}^{2}}{2\omega^{2}_{ci}}
\left\langle\left(\mathbf{k}\left(t\right)\left(\delta\phi\left(t\right)
-\delta\phi\left(t_{1}\right)\right)\right)^{2}\right\rangle=2t^{2}\int\limits^{t}_{t_{1}}d\hat{t}
\frac{C\left(\mathbf{k},\hat{t}\right)}{\hat{t}^{2}}.\label{61}
\end{eqnarray}

\section*{V.NONLINEAR NON-MODAL EVOLUTION \\ OF THE KINETIC DRIFT INSTABILITY\\ IN SHEAR FLOW}

Now we use Eq.(\ref{61}) with $v_{\perp}^{2}$ changed on
$v_{Ti}^{2}$ in Eq.(\ref{50}) and integrate (\ref{50}) over
$v_{\perp}$. We obtain in ion terms additional multiplier, $\exp
(-t^{2}\int\limits^{t}_{t_{1}}d\hat{t}\hat{t}^{-2}
C\left(\mathbf{k},\hat{t}\right))$, as compared with linear
non-renormalised equation(\ref{22}),
\begin{eqnarray}
&\displaystyle k^{2}\left(t\right)\varphi\left(\mathbf{k},t\right)=
\frac{i}{\lambda^{2}_{Di}}\int\limits^{t}_{t_{0}}dt_{1}
\varphi\left(\mathbf{k},t_{1}\right)I_{0}\left(k_{\perp}\left(t\right)k_{\perp}\left(t_{1}\right)
\rho^{2}_{i}\right)e^{-t^{2}\int\limits^{t}_{t_{1}}d\hat{t}
\frac{C\left(\mathbf{k},\hat{t}\right)}{\hat{t}^{2}}}
e^{-\frac{1}{2}\rho^{2}_{i}\left(k^{2}_{\perp}\left(t\right)
+k^{2}_{\perp}\left(t_{1}\right)\right)-\frac{1}{2}k^{2}_{z}v^{2}_{Ti}\left(t-t_{1}\right)^{2}}
\nonumber \\
&\displaystyle \times\left[k_{y}v_{di}
+ik^{2}_{z}v^{2}_{Ti}\left(t-t_{1}\right)\right]+\frac{i}{\lambda^{2}_{De}}\int\limits^{t}_{t_{0}}dt_{1}
\varphi\left(\mathbf{k},t_{1}\right)I_{0}\left(k_{\perp}\left(t\right)k_{\perp}\left(t_{1}\right)
\rho^{2}_{e}\right)
e^{-\frac{1}{2}\rho^{2}_{i}\left(k^{2}_{\perp}\left(t\right)
+k^{2}_{\perp}\left(t_{1}\right)\right)-\frac{1}{2}k^{2}_{z}v^{2}_{Ti}\left(t-t_{1}\right)^{2}}
\nonumber \\
&\displaystyle \times\left[k_{y}v_{de}
+ik^{2}_{z}v^{2}_{Te}\left(t-t_{1}\right)\right]-4\pi\sum_{\alpha=i,e}e_{\alpha}\delta
n_{\alpha}\left(\mathbf{k}, t, t_{0}\right).\label{62}
\end{eqnarray}
For the perturbations of drift type, the equation for
$\Phi\left(\mathbf{k},t\right)=
\varphi\left(\mathbf{k},t\right)\Theta \left(\mathbf{k},t\right)$
has a form
\begin{eqnarray}
&\displaystyle \int\limits^{t}_{t_{0}}dt_{1}\left\lbrace
\frac{d}{dt_{1}}\Phi\left(\mathbf{k},t_{1}\right) \left(1+\tau
\right)-\left[\frac{d}{dt_{1}}\left(\Phi\left(\mathbf{k},t_{1}\right)
I_{0}\left(k_{\perp}\left(t\right)k_{\perp}\left(t_{1}\right)\rho^{2}_{i}\right)
e^{-\frac{1}{2}\rho^{2}_{i}\left(k^{2}_{\perp}\left(t\right)+k^{2}_{\perp}
\left(t_{1}\right)\right)}\right)\right.\right.
\nonumber \\
&\displaystyle
\left.\left.-\frac{d}{dt_{1}}\left(\Phi\left(\mathbf{k},t_{1}\right)\left(1-e^{-t^{2}\int\limits^{t}_{t_{1}}d\hat{t}
\frac{C\left(\mathbf{k},\hat{t}\right)}{\hat{t}^{2}}}\right)
I_{0}\left(k_{\perp}\left(t\right)k_{\perp}\left(t_{1}\right)\rho^{2}_{i}\right)
e^{-\frac{1}{2}\rho^{2}_{i}\left(k^{2}_{\perp}\left(t\right)+k^{2}_{\perp}
\left(t_{1}\right)\right)}\right)
\right]e^{-\frac{1}{2}k^{2}_{z}v^{2}_{Ti}
\left(t-t_{1}\right)^{2}}\right\rbrace \nonumber
\\ &\displaystyle
=i\int\limits^{t}_{t_{0}}dt_{1}
\Phi\left(\mathbf{k},t_{1}\right)k_{y}v_{di}
I_{0}\left(k_{\perp}\left(t\right)k_{\perp}\left(t_{1}\right)\rho^{2}_{i}\right)
e^{-\frac{1}{2}\rho^{2}_{i}\left(k^{2}_{\perp}\left(t\right)
+k^{2}_{\perp}\left(t_{1}\right)\right) } \nonumber
\\ &\displaystyle-i\int\limits^{t}_{t_{0}}dt_{1}
\Phi\left(\mathbf{k},t_{1}\right)k_{y}v_{di}
I_{0}\left(k_{\perp}\left(t\right)k_{\perp}\left(t_{1}\right)\rho^{2}_{i}\right)
e^{-\frac{1}{2}\rho^{2}_{i}\left(k^{2}_{\perp}\left(t\right)
+k^{2}_{\perp}\left(t_{1}\right)\right)}\left(1-e^{-t^{2}\int\limits^{t}_{t_{1}}d\hat{t}
\frac{C\left(\mathbf{k},\hat{t}\right)}{\hat{t}^{2}}}\right)+\nonumber
\\ &\displaystyle +\tau\int\limits^{t}_{t_{0}}dt_{1}
\left(\frac{d\Phi\left(\mathbf{k},t_{1}\right)}{dt_{1}}
+ik_{y}v_{de}\Phi\left(\mathbf{k},t_{1}\right)\right)e^{-\frac{1}{2}k^{2}_{z}v^{2}_{Te}
\left(t-t_{1}\right)^{2}}.\label{63}
\end{eqnarray}
If $k_{\perp}\rho_{i}<1$ at time $t=0$ (at which the shear flow
emerge), we will get $k_{\perp}\left(t\right)\rho_{i}<1$ on times
$t<t_{s}$ throughout. By using the approximation
\begin{eqnarray}
&\displaystyle
I_{0}\left(k_{\perp}\left(t\right)k_{\perp}\left(t_{1}\right)\rho^{2}_{i}\right)
e^{-\frac{1}{2}\rho^{2}_{i}\left(k^{2}_{\perp}\left(t\right)
+k^{2}_{\perp}\left(t_{1}\right)\right)} \nonumber
\\ &\displaystyle\approx
b_{i}+\left(k_{x}k_{y}V'_{0}\left(t+t_{1}\right)-\frac{1}{2}k^{2}_{y}
\left(V'_{0}\right)^{2}\left(t^{2}+t_{1}^{2}\right)\right)\rho^{2}_{i}\Theta\left(t\right),\label{64}
\end{eqnarray}
where $b_{i}=1-k^{2}_{\perp}\rho^{2}_{i}$, and
$\Theta\left(t\right)$ indicates that the shear flow emerge at
$t=0$, we present Eq.(\ref{63}) in the form
\begin{eqnarray}
&\displaystyle \int\limits^{t}_{t_{0}}dt_{1} \left(\frac{d
\Phi\left(\mathbf{k},t_{1}\right)}{dt_{1}}+i\omega\left(\mathbf{k}\right)
\Phi\left(\mathbf{k},t_{1}\right)\right)\nonumber
\\ &\displaystyle=-\frac{b_{i}}{a_{i}}\int\limits^{t}_{t_{0}}dt_{1}\left(\frac{d
\Phi\left(\mathbf{k},t_{1}\right)}{dt_{1}}+ik_{y}v_{di}
\Phi\left(\mathbf{k},t_{1}\right)\right)\left(1-e^{-\frac{1}{2}k^{2}_{z}v^{2}_{Ti}
\left(t-t_{0}\right)^{2}}\right)\nonumber
\\ &\displaystyle-\frac{b_{i}}{a_{i}}\int\limits^{t}_{t_{0}}dt_{1}\left(\frac{d
\Phi\left(\mathbf{k},t_{1}\right)}{dt_{1}}+ik_{y}v_{di}
\Phi\left(\mathbf{k},t_{1}\right)\right)\left(1-\exp\left[ -t^{2}\int\limits^{t}_{t_{1}}d\hat{t}\:
\frac{C\left(\mathbf{k},\hat{t}\right)}{\hat{t}^{2}}\right] \right)\nonumber
\end{eqnarray}
\begin{eqnarray} &\displaystyle+\int\limits^{t}_{0}dt_{1}\left(\frac{d
\Phi\left(\mathbf{k},t_{1}\right)}{dt_{1}}+ik_{y}v_{di}
\Phi\left(\mathbf{k},t_{1}\right)\right)\left(\frac{k_{x}}{k_{y}}\frac{\left(t+t_{1}\right)}
{a_{i}V'_{0}t^{2}_{s}}-\frac{\left(t^{2}+t^{2}_{1}\right)}{2a_{i}t^{2}_{s}}\right)\nonumber \\
&\displaystyle+\int\limits^{t}_{0}dt_{1}\Phi\left(\mathbf{k},t_{1}\right)
\frac{1}{a_{i}V'_{0}t^{2}_{s}}\left(\frac{k_{x}}{k_{y}}-V'_{0}t\right)\nonumber \\
&\displaystyle+\frac{\tau}{a_{i}}\int\limits^{t}_{t_{0}}dt_{1}\left(\frac{d
\Phi\left(\mathbf{k},t_{1}\right)}{dt_{1}}+ik_{y}v_{de}
\Phi\left(\mathbf{k},t_{1}\right)\right)\left(1-e^{-\frac{1}{2}k^{2}_{z}v^{2}_{Te}
\left(t-t_{0}\right)^{2}}\right),\label{65}
\end{eqnarray}
where $\omega\left(\mathbf{k}\right)$ is determined by
Eq.(\ref{28}). The right hand side of Eq.(\ref{65}) is small for
$\left(V_{0}'\right)^{-1}<t<t_{s}$, for $\tau<1$ and for weak ion
Landau damping. Therefore, the solution to Eq.(\ref{65}) we seek in
the form
\begin{eqnarray}
&\displaystyle
\Phi\left(\mathbf{k},t\right)=C\exp\left(-i\omega\left(\mathbf{k}\right)t+i\nu
\left(\mathbf{k}, t\right)\right).\label{66}
\end{eqnarray}
Applying the procedure of the solution of the integral equation
(\ref{25}) to the renormalized version of that equation, (\ref{64}),
we obtain for $\left(V'_{0}\right)^{-1}<t< t_{s}$ the solution to
Eq.(\ref{65}) in the form
\begin{eqnarray}
&\displaystyle \Phi\left(\mathbf{k},t\right)=\Phi_{0}\exp
\left[-i\omega\left(\mathbf{k}\right)t\left(1-\frac{1+\tau}{a_{i}b_{i}}
\frac{t^{2}}{3t_{s}^{2}}\right)t+iRe
\delta\omega\left(\mathbf{k}\right)t\right.\nonumber
\\ &\displaystyle\left.+\left(\gamma\left(\mathbf{k}\right)
-\frac{t}{2a_{i}t_{s}^{2}}\right)t-
\int\limits^{t}_{0}C\left(\mathbf{k}, t_{1}\right)dt_{1}\right],
\label{67}
\end{eqnarray}
where $C\left(\mathbf{k}, t\right)$ is determined by the equation
\begin{eqnarray}
&\displaystyle C\left(\mathbf{k},
t\right)=\frac{c^{2}}{B^{2}}k_{y}^{2}\rho^{2}_{i}\frac{\left(V'_{0}t\right)^{6}}{8}\int
d\mathbf{k}_{1}\left|\varphi\left(\mathbf{k}_{1},
t\right)\right|^{2}C\left(\mathbf{k}_{1},
t\right)\frac{k_{1y}^{4}}{\omega^{2}\left(\mathbf{k}_{1}\right)}.
\label{68}
\end{eqnarray}
If we omit linear non-modal terms in Eq.(\ref{67}), the condition of
the balance of the linear modal growth of the kinetic drift
instability and non-linear non-modal dumping is determined by the
equation $\gamma \left(\mathbf{k}\right)=C\left(\mathbf{k},
t\right)$. By using this equation in Eq.(\ref{68}), we obtain the
equation, which determines the time, at which that balance occurs,
\begin{eqnarray}
&\displaystyle
\frac{\gamma\left(\mathbf{k}\right)}{\left(V'_{0}t\right)^{6}}
=\frac{c^{2}}{8B^{2}}k_{y}^{2}\rho^{2}_{i}\int
d\mathbf{k}_{1}\left|\varphi\left(\mathbf{k}_{1},
t\right)\right|^{2}\gamma\left(\mathbf{k}_{1}\right)
\frac{k_{1y}^{4}}{\omega^{2}\left(\mathbf{k}_{1}\right)}. \label{69}
\end{eqnarray}
The effect of the shear flow reveals in the reducing with time as
$\left(V'_{0}t\right)^{-6}$ the magnitude of the growth rate in the
left part of the balance equation (\ref{69}). That causes rapid
suppression of the drift turbulence. The evolution of drift
turbulence in times $t\geq t_{s} $ continues as strongly non-modal
process, for which Markovian approximation, which is admissible for
the solution Eq.(\ref{23}) with small growth rate and non-modal
terms with respect to the frequency $\omega
\left(\mathbf{k}\right)$, ceases to be valid.

\section*{VI.CONCLUSIONS}
In this paper, by using method of shearing modes or non-modal
approach, originally developed by Lord Kelvin\cite{Kelvin} for fluid
descriptions of fluid shear flows, we develop for the first time
non-modal kinetic theory of plasma shear flow directed across the
magnetic field. We obtain linear, (\ref{22}), and renormalized
nonlinear, (\ref{62}), governing integral equations for the
perturbed electrostatic potential. By using these equations we
obtain linear, (\ref{32}), (\ref{41}), and renormalized nonlinear,
(\ref{67}), initial value problems solutions for kinetic drift
instability of plasma shear flow. Obtained solutions display two
distinct non-modal effects, which are observed in laboratory frame
of reference.

The first effect displays the inhomogeneous Doppler shift, which is
presented by the term $V'_{0}tk_{y}x$ in exponential of
Eq.(\ref{33}). This term displays the shearing of waves patterns in
shear flow, observed in laboratory frame; in time
$t>\left(k_{y}V'_{0}/k_{x}\right)^{-1}$, turbulence becomes almost
one-dimensional in plane across the magnetic field and directed
almost along the shear flow. It is obvious that this Doppler shift
is irrelevant to the suppression of drift turbulence by shear flow.

Second non-modal effect is of principal importance for turbulence
evolution in plasma shear flows. It reveals as a time dependent
finite Larmor radius term in governed equations (\ref{22}) and
(\ref{62}). The time dependence originates from the dot product of
the time dependent coordinates of ion gyration in sheared
coordinates, (Eqs.(\ref{12})), and wave number, which is time
independent in these coordinates. That term completely conserves its
form after the transformation to the laboratory frame variables, in
which ion Larmor orbit is almost circular in shear flow with
$\left|V'_{0}\right|\ll\omega_{ci}$ \cite{Shaing}. In laboratory
frame of references, this non-modal effect is seemed as resulted
from the coupling of the ion gyration and temporal variation of the
wave number,
$k_{\perp\left(lab\right)}=\left(k_{y}^{2}+\left(k_{x}-k_{y}V'_{0}t\right)^{2}\right)^{1/2}$
of the shearing mode. Linear theory reveals (see
Eqs.(\ref{32}),(\ref{41})) that on the times
$\left(V'_{0}\right)^{-1}<t<\left(V'_{0}k_{y}\rho_{i}\right)^{-1}$
this time dependence resulted in the reducing the frequency and
growth rate of the drift kinetic instability and to gradual
suppression of the instability. It is important to note, that this
effect is absent in the theory grounded on drift kinetic equation,
in which effects of the finite Larmor radius are ignored. It is
interesting to note that similar effect of the non--modal evolution
of the perturbed electrostatic potential, which consists in reducing
with time the frequency and growth rate, was
discovered\cite{Mikhailenko-2000} in the investigations of the
resistive drift instability on the base of the Hasegawa-Wakatani
system. That effect for resistive drift instability originates from
the time dependent polarization drift which is in fact the
manifestation of the time dependent effect of the finite ion Larmor
radius.

Decisive impact on the temporal evolution of the kinetic drift
instability has non-linear non--modal effect of the turbulent
scattering of ions by the ensemble of sheared waves. We find that
turbulent scattering of the gyrophase of ion Larmor orbit is the
dominant effect, which determines extremely rapid suppression of
drift turbulence by flow shear.

\section*{Acknowledgements}

Authors acknowledge useful conversations with S.I.Krasheninnikov. We
are also grateful the Erasmus Mundus Foundation for partial
financial support this research.

\section*{Appendix 1.Equilibrium distribution function $F_{0\alpha}$.}
The equilibrium distribution function $F_{0\alpha}\left(x, v_{x},
v_{y}, v_{z}\right)$ is governed by the equation
\begin{equation*}
v_{x}\frac{\partial F_{0\alpha}}{\partial
x}+\left(\frac{e_{\alpha}}{m_{\alpha}}E_{0}\left(x\right)
+\omega_{c\alpha}v_{y} \right)\frac{\partial F_{0\alpha}}{\partial
v_{x}}-\omega_{c\alpha}v_{x}\frac{\partial F_{0\alpha}}{\partial
v_{y}}=0.\eqno(A.1)
\end{equation*}
From characteristic equations
\begin{equation*}
\frac{dx}{v_{x}}=\frac{dv_{x}}{\dfrac{e}{m}E_{0}\left(x\right)+\omega_{c\alpha}v_{y}}
=-\frac{dv_{y}}{\omega_{c\alpha}v_{x}}\eqno(A.2)
\end{equation*}
the relation
\begin{equation*}
-\frac{e}{m}\omega_{c}E_{0}\left(x\right)dx+\omega_{c\alpha}v_{y}dv_{y}
+\omega_{c\alpha}v_{x}dv_{x}=d\left(\omega_{c}\textit{H}\right)=0\eqno(A.3)
\end{equation*}
follows in which $\textit{H}$ is Hamiltonian of a particle  in
electric and magnetic fields. By using the expansions
\begin{equation*}
E_{0}\left(x\right)=E_{0}\left(X_{\alpha}\right)+E'_{0}
\left(X_{\alpha}\right)\left(x-X_{\alpha}\right),\eqno(A.4)
\end{equation*}
and
\begin{equation*}
v_{y}=v_{y\alpha}+V_{0}\left(X_{\alpha}\right)
+V'_{0}\left(X_{\alpha}\right)\left(x-X_{\alpha}\right),\eqno(A.5)
\end{equation*}
and accounting for that
$V_{0}\left(X_{\alpha}\right)=-cE_{0}\left(X_{\alpha}\right)/B_{0}$
and $dv_{y}=-\omega_{c\alpha}dx$, we obtain that
\begin{equation*}
-\omega^{2}_{c\alpha}v_{\alpha
y}dx_{\alpha}+\omega_{c\alpha}v_{x\alpha}dv_{x\alpha}=d\left(\omega_{c}\textit{H}\right)=0,\eqno(A.6)
\end{equation*}
where identities $x=x_{\alpha}$ and $v_{x}=v_{x\alpha}$ were used.
It follows from (\textit{A}.5) that
$dv_{y}=dv_{y\alpha}+V'_{0}\left(X_{\alpha}\right)dx_{\alpha}$.
Therefore
$dx_{\alpha}=dx=-dv_{y}/\omega_{c\alpha}=-\left(dv_{y\alpha}+V'_{0}\left(X_{\alpha}\right)
dx_{\alpha}\right)/\omega_{c\alpha}$ and
$dx_{\alpha}=-dv_{y\alpha}/\mu\omega_{c\alpha}$. With Eqs.(\ref{13})
it follows that
\begin{equation*}
\frac{1}{2}d\left(\omega_{c\alpha}v_{\perp}^{2}\right)=d\left(\omega_{c}\textit{H}\right)=0.\eqno(A.7)
\end{equation*}
The same conclusion about absence of the spatial dependence in
Hamiltonian in sheared coordinates follows from the analysis of the
characteristics of Eq.(\ref{8}).

\end{document}